\documentclass[12pt]{article}

\usepackage{srcltx}
\usepackage[dvips,letterpaper]{geometry}
\usepackage{graphicx,psfrag,epsf}
\usepackage{times}
\usepackage{fullpage}
\usepackage{setspace}
\usepackage{amsmath,amssymb,amsfonts}
\usepackage{bm}
\usepackage{graphicx}
\usepackage{epsfig}
\usepackage{here}
\usepackage{subfigure}
\usepackage{url}
\usepackage{verbatim}
\usepackage{natbib}
\usepackage{color}
\usepackage{booktabs}
\usepackage{graphicx,psfrag,epsf}
\usepackage{lscape}
\usepackage{placeins}

\input cyracc.def

\title{Parametric quantile regression models for fitting double bounded response with application to COVID-19 mortality rate data}

\author{\normalsize
\textbf{Diego I. Gallardo}$^{1}$\thanks{Corresponding
author: Diego I. Gallardo. Departamento de Matem\'atica, Facultad de Ingenier\'ia, Universidad de Atacama, Copiap\'o, Chile. Email: diego.gallardo@uda.cl}, \quad \textbf{Marcelo Bourguignon}$^{2}$, \\ \normalsize \textbf{Yolanda M. G\'omez}$^{1}$ \quad \textbf{and} \quad \textbf{Christian Caama\~no-Carrillo}$^{3}$
\\
{\footnotesize $^{1}$Departamento de Matem\'atica, Facultad de Ingenier\'ia, Universidad de Atacama, Copiap\'o, Chile}\\[-0.15cm]
{\footnotesize $^{2}$Departamento de Estat\'istica, Universidade Federal do Rio Grande do Norte, Natal, Brazil}\\[-0.15cm]
{\footnotesize $^{3}$Departamento de Estad\'istica, Facultad de Ciencias, Universidad del B\'io-B\'io, Concepci\'on, Chile}\\[-0.15cm]
}

\date{}

\begin{document}
%\linenumbers

\maketitle

\vspace{-0.7cm}
\begin{abstract}
In this paper, we develop two fully parametric quantile regression models, based
on power Johnson $S_B$ distribution Cancho et al. [Statistical Methods in Medical Research, 2020], for modeling
unit interval response at different quantiles. In particular, the conditional distribution is modelled by the power Johnson SB distribution. The maximum likelihood
method is employed to estimate the model parameters.
Simulation studies are conducted to evaluate the performance of the maximum likelihood estimators in finite samples.
Furthermore, we discuss residuals and influence diagnostic tools.
The effectiveness of our proposals is illustrated with two data set given by the mortality rate of COVID-19 in different countries.

\vspace{-0.3cm}

\end{abstract}

\vspace{-0.3cm}

\section{Introduction}\label{sec:1}
\noindent

The most commonly employed two-parameter distribution for modeling
doubly bounded random variables on the unit interval is the beta distribution.
In order to accommodate explanatory variable in the modeling, \cite{ferrari04} introduced the beta regression model
based on a parameterization of the beta distribution in terms of the mean and precision parameters.
A substantial number of practical and theoretical works have focused on the use of the mean
reparameterized beta distribution as an integral of the model.
For example, see \cite{ospina08}, \cite{Bayes12} and \cite{migri18}.
However, there are limitations of the conditional mean models.
For example, in an assymetric distribution, or in the presence of outliers, the mean
is pulled in the direction of the tail, making it a less representative measure of central tendency.

Quantile regression, introduced by \cite{Koenker1978}, is a methodology for understanding the
conditional distribution of a response variable given the values of some covariates at different levels (quantiles), thus providing users with a more complete picture. In particular, several authors \citep{su15, Lemonte20} highlighted the robustness to outliers connected with quantile regression models. Furthermore, if the conditional dependent variable is skewed, the quantiles may be more appropriate when compared with the mean \citep{mazu20}.

However, parametric quantile regression models for limited range response variables
has not received much attention in the literature.
\citet{Lemonte16} introduced a new class of distributions named the generalized Johnson $S_B$
with bounded support on the basis of the symmetric family of distributions.
In particular, \citet{Lemonte16} provided the median re-parameterizations of the Johnson $S_B$
distribution \citep{Johnson} that facilitates its use in a regression setting.
Unlike the beta regression, the median in the re-parameterized Johnson $S_B$ distribution is related to a linear predictor.
\cite{Cancho} generalized the Johnson $S_B$ model to a general class of
distributions. The authors introduced an extra parameter to model the shape of the Johnson $S_B$ distribution, and
studied a quantile regression model for limited range response variables.
However, they consider the model only based on the normal distribution.
Other quantile regression models for limited range response
variables are presented in \cite{Bayes17}, \cite{mazu20} and \cite{Lemonte20}.

In this paper, we formulate %in Section \ref{sec:2}
two rich classes of parametric quantile regression models for a bounded response, where the response
variable is power Johnson $S_B$ distributed \citep{Cancho} using a new parametrization of this
distribution that is indexed by quantile (not only for median regression) and shape parameters.
The estimation and inference for the proposed quantile regression models can be carried out based on the likelihood paradigm
(parametric approach). Also, we give full diagnostic tools for detecting possible outliers and discuss a type of residuals.
The main motivations for these new parametric quantile regression models are fourfold:
(i) the Johnson $S_B$ and power Johnson $S_B$ regression models are themselves special
cases of the proposed quantile models;
(ii) the first proposed model has a parameter which controls the shape
and skewness of the distribution;
(iii) the second proposed model has less computational cost; and
(iv) we considered the model based on several models (logistic, Cauchy and normal) and several link functions.

The article is organized as follows. In Section \ref{sec:2}, we construct two new quantile regression models for
bounded response variables. Estimation, residuals and diagnostic measures are discussed in Section \ref{sec:3}.
Section \ref{sec:4} discusses some simulation results for the maximum likelihood (ML) estimation method.
The effectiveness of our models is illustrated in Section \ref{sec:5} by using the mortality rate of COVID-19 in different countries. Final comments are presented in Section \ref{sec:6}. This paper contains an additional application related to the reproductive activity of the anchoveta in Chile in a Supplementary Material.

\section{The generalized Johnson $S_B$ distribution}\label{sec:2}
\noindent

\citet{Lemonte16} introduced a new class of distributions named the generalized Johnson $S_B$ (``GJS'' for short) distribution.
The class is defined by the transformation $Y=Q^{-1}((X-\gamma)/\delta) \in (0,1)$, where $\gamma \in \mathbb{R}$, $\delta>0$, $Q(y)=\log(y/(1-y))$ is the logit function (also representing the quantile function for the standard logistic distribution) and $X\sim S(0,1; g)$, i.e., the symmetrical family of distributions with pdf given by $g(w)$, $w\in \mathbb{R}$, where $g$ is a function such as $g: \mathbb{R}\rightarrow [0,\infty)$. Considering the reparametrization $\gamma=-\delta Q(\xi)$, the cdf of the GJS is given by
\begin{equation*}
F(y;\xi,\delta)=\int_{-\infty}^{\delta[Q(y)-Q(\xi)]} g(u)\textrm{d}u, \quad y, \xi \in (0,1).
\end{equation*}
As $F(\xi;\xi,\delta)=1/2$, the parameter $\xi$ represents directly the median of the distribution. Additionally, the authors interpret $\delta$ as a dispersion parameter. Therefore, a regression structure on $\xi$ and $\delta$ is studied by the authors, providing a rich class of median regression model with varying dispersion. \cite{Cancho} considered $g(u)=\phi(u)$ (where $\phi(\cdot)$ denotes the pdf of the standard normal model) and the power model transformation \citep{Lehmann, Durrans} to extend this class of models (named as PJSB), which cdf is given by
\begin{equation*}
F(y;\alpha,\gamma,\delta)=[\Phi(\gamma+\delta Q(y))]^\alpha, \quad y \in (0,1), \alpha,\delta>0, \gamma \in \mathbb{R}.
\end{equation*}
Besides the logistic model, the authors also considers $Q(y)$ as the quantile function for the normal, Cauchy, Gumbel and reverse Gumbel models. Thus, the pdf of the PJSB model is
\begin{equation*}
f(y;\gamma,\delta,\alpha)=\delta\alpha[\Phi(\gamma+\delta Q(y))]^{\alpha-1}\phi(\gamma+\delta Q(y))\left|\frac{\textrm{d} Q(y)}{\textrm{d} y}\right|, \quad y \in (0,1).
\end{equation*}
Defining $x_q=\Phi^{-1}(q^{1/\alpha})$, the authors considered the reparametrization $\psi=Q^{-1}\left(\frac{x_{0.5}(\alpha)-\gamma}{\delta}\right)$, which represents the median of the PJSB distribution (for any $Q(\cdot)$ quantile function). As $\gamma=x_{0.5}(\alpha)-\delta Q(\psi)$, the pdf of the PJSB can be expressed as
\begin{equation*}
f(y;\psi,\delta,\alpha)=\delta\alpha[\Phi(\delta [Q(y)-Q(\psi)]+x_{0.5}(\alpha))]^{\alpha-1}\phi(\delta [Q(y)-Q(\psi)+x_{0.5}(\alpha)]\left|\frac{\textrm{d} Q(y)}{\textrm{d} y}\right|, \quad y \in (0,1).
\end{equation*}
The authors proposed a regression model for $\psi$ and $\delta$ in this model. However, this model can be restrictive because considers the only normal distribution. For this reason, we consider the power model transformation of \citet{Lehmann, Durrans} for the GJS distribution of \cite{Lemonte16}, say the power generalized Johnson $S_B$ (PGJSB) distribution, with cdf given by
\begin{equation}
F(y;\xi,\delta,\alpha)=\left(\int_{-\infty}^{\delta[Q(y)-Q(\xi)]} g(u)\textrm{d} u\right)^\alpha=[G(\delta [Q(y)-Q(\xi)])]^\alpha=[G(\gamma+\delta Q(y))]^\alpha, \quad y \in (0,1) \label{cdf.PGJS},
\end{equation}
and pdf given by
\begin{equation*}
f(y;\gamma,\delta,\alpha)=\delta\alpha[G(\gamma+\delta Q(y))]^{\alpha-1}g(\gamma+\delta Q(y))\left|\frac{\textrm{d} Q(y)}{\textrm{d} y}\right|, \quad y \in (0,1).\label{model1}
\end{equation*}
where $G$ is the cdf related to $g$. Evidently, for $G=\Phi$, we recover the model in \cite{Cancho}. However, we are interested in model a general quantile, say $q$, not only the median. In this work, we discuss two ways to model the $100\times q$th quantile considering the PGJSB model.

\begin{enumerate}
\item We note that $\psi=Q^{-1}\left(\frac{x^*_{q}(\alpha)-\gamma}{\delta}\right)$ is the $100\times q$th quantile for the PGJSB model, where $x^*_q(\alpha)=G^{-1}(q^{1/\alpha})$. Based on this idea, we also can reparametrize the model noting defining $\gamma=x^*_q(\alpha)-\delta Q(\psi)$. The pdf for this reparametrization is
\begin{equation}
f(y;\psi,\delta,\alpha)=\delta\alpha[G(\delta [Q(y)-Q(\psi)]+x^*_{q}(\alpha))]^{\alpha-1}g(\delta [Q(y)-Q(\psi)+x^*_{q}(\alpha)]\left|\frac{\textrm{d} Q(y)}{\textrm{d} y}\right|, \quad y \in (0,1).\label{pdf.RPGJSB1}
\end{equation}
In this work, we will refers to this specific parametrization as RPGJSB1$_q(\psi,\delta,\alpha)$.
\item Despite the nature of $\alpha$ is to be a parameter, we can consider $\alpha(q)=-\log(q)/\log(2)$, $q\in (0,1)$ as fixed. With this definition, the cdf in (\ref{cdf.PGJS}) evaluated in $\xi$ is given by $F(\xi;\xi,\delta)=(1/2)^{\alpha(q)}=q$. Therefore, fixing $\alpha(q)=-\log(q)/\log(2)$, $q\in (0,1)$, we have that $\xi$ represents the $100 \times q$th quantile of the distribution and similarly to the work of \citet{Lemonte16}, $\delta$ also can be interpreted as a dispersion parameter. We will refers to this parametrization as RPGJSB2$_q(\xi,\delta)$.
\end{enumerate}

\begin{figure}[!htbp]
\centering
\psfrag{0}[c][c]{\scriptsize{0}}
\psfrag{1}[c][c]{\scriptsize{1}}
\psfrag{2}[c][c]{\scriptsize{2}}
\psfrag{3}[c][c]{\scriptsize{3}}
\psfrag{4}[c][c]{\scriptsize{4}}
\psfrag{5}[c][c]{\scriptsize{5}}
\psfrag{6}[c][c]{\scriptsize{6}}
\psfrag{7}[c][c]{\scriptsize{7}}
\psfrag{0.0}[c][c]{\scriptsize{0.0}}
\psfrag{0.2}[c][c]{\scriptsize{0.2}}
\psfrag{0.4}[c][c]{\scriptsize{0.4}}
\psfrag{0.6}[c][c]{\scriptsize{0.6}}
\psfrag{0.8}[c][c]{\scriptsize{0.8}}
\psfrag{1.0}[c][c]{\scriptsize{1.0}}
\psfrag{AxB}[c][c]{\scriptsize{$f(y)$}}
\psfrag{AxA}[c][c]{\scriptsize{$y$}}
\psfrag{BxB}[c][c]{\hspace{6mm} \scriptsize{$\psi=0.10$}}
\psfrag{BxC}[c][c]{\hspace{6mm} \scriptsize{$\psi=0.25$}}
\psfrag{BxD}[c][c]{\hspace{6mm} \scriptsize{$\psi=0.50$}}
\psfrag{BxE}[c][c]{\hspace{6mm} \scriptsize{$\psi=0.75$}}
\psfrag{BxF}[c][c]{\hspace{6mm} \scriptsize{$\psi=0.90$}}
{\includegraphics[height=5cm,width=5cm,angle=0]{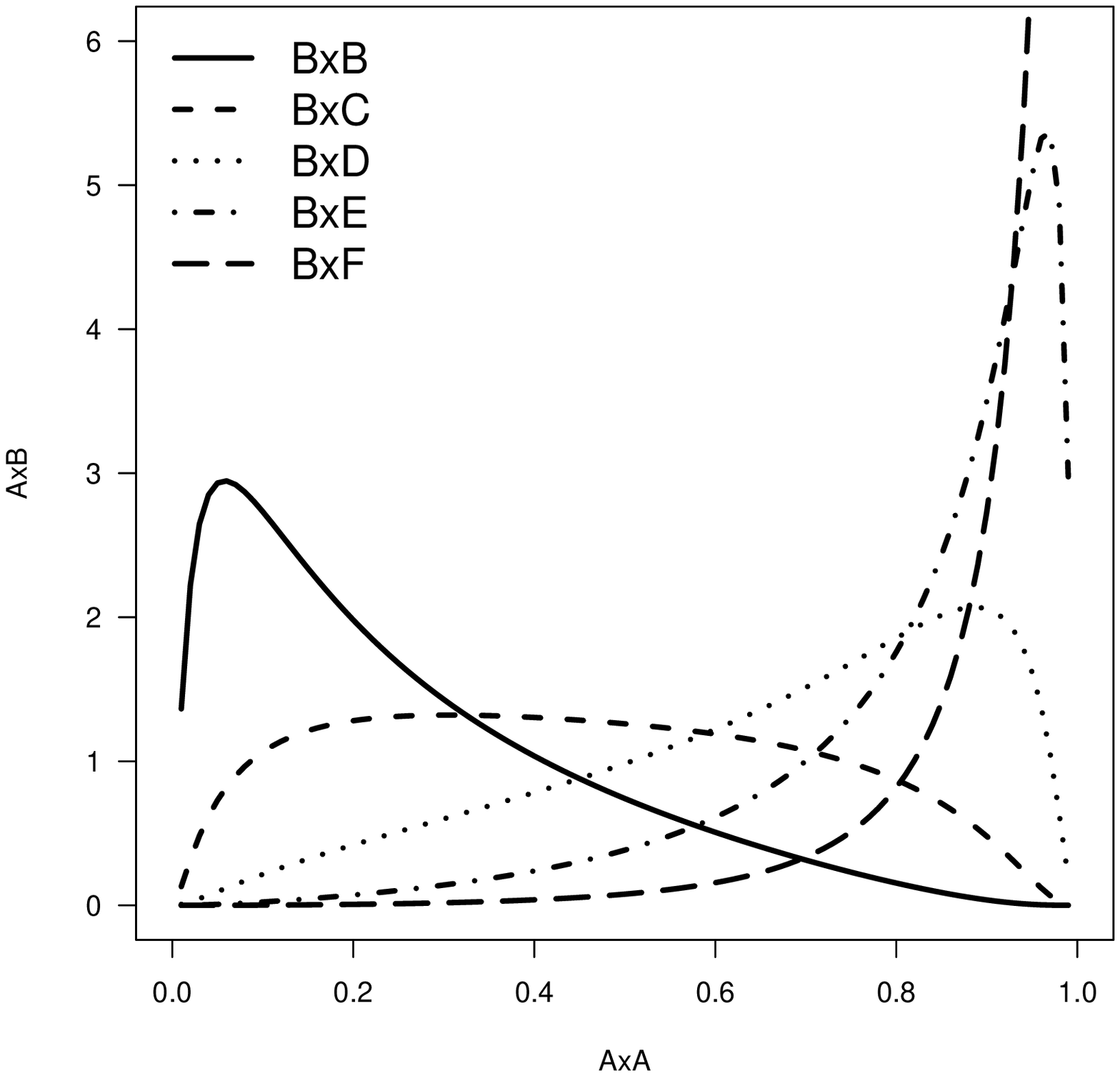}}
\psfrag{BxB}[c][c]{\hspace{6mm}\scriptsize{$\psi=0.10$}}
\psfrag{BxC}[c][c]{\hspace{6mm}\scriptsize{$\psi=0.25$}}
\psfrag{BxD}[c][c]{\hspace{6mm}\scriptsize{$\psi=0.50$}}
\psfrag{BxE}[c][c]{\hspace{6mm}\scriptsize{$\psi=0.75$}}
\psfrag{BxF}[c][c]{\hspace{6mm}\scriptsize{$\psi=0.90$}}
{\includegraphics[height=5cm,width=5cm,angle=0]{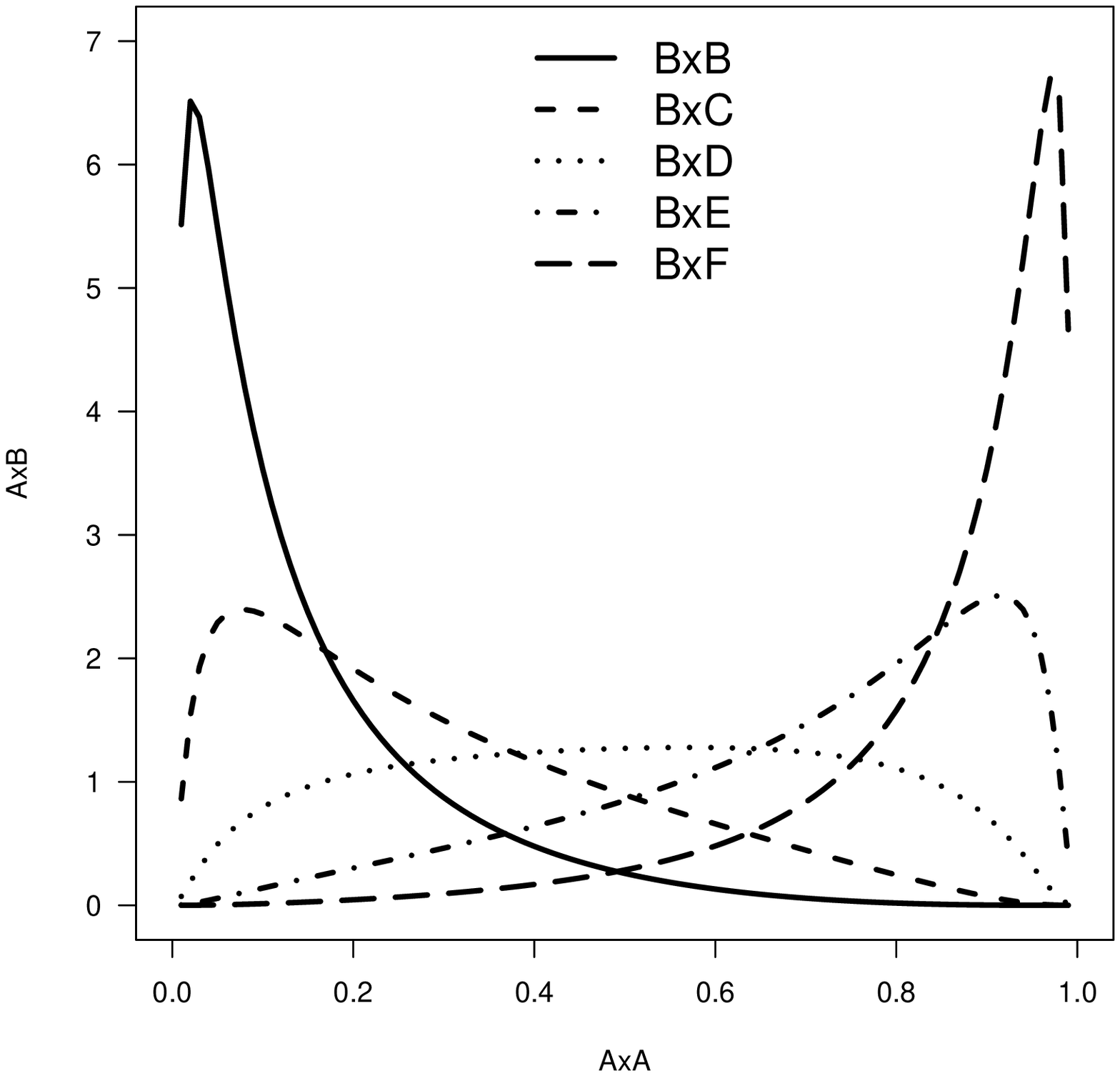}}
\psfrag{BxB}[c][c]{\hspace{6mm}\scriptsize{$\alpha=0.10$}}
\psfrag{BxC}[c][c]{\hspace{6mm}\scriptsize{$\alpha=0.25$}}
\psfrag{BxD}[c][c]{\hspace{6mm}\scriptsize{$\alpha=0.50$}}
\psfrag{BxE}[c][c]{\hspace{6mm}\scriptsize{$\alpha=1.00$}}
\psfrag{BxF}[c][c]{\hspace{6mm}\scriptsize{$\alpha=5.00$}}
{\includegraphics[height=5cm,width=5cm,angle=0]{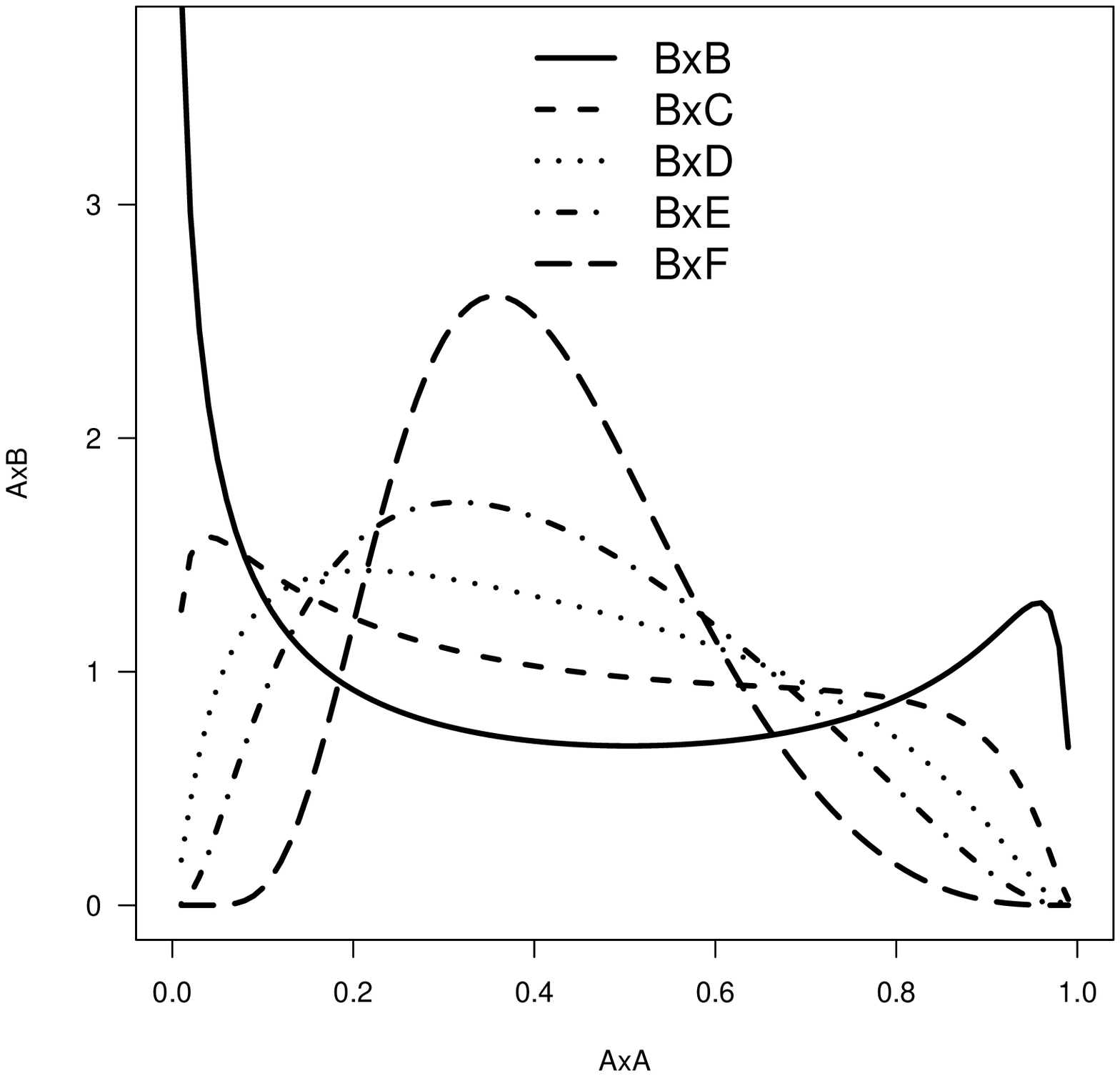}}
\caption{Pdf for RPGJSB1$_q(\psi,\delta=1,\alpha)$ model with logit link and $G=\Phi$. Left panel: $q=0.25$, $\alpha=0.5$ and varying $\psi$; center panel: $q=0.5$, $\alpha=0.5$ and varying $\psi$; right panel: $q=0.5$, $\psi=0.4$ and varying $\alpha$.}
 \label{d.RPGJSB1}
\end{figure}

In both cases, the RPGJSB1$_q$ and RPGJSB2$_q$ models can be used to define a rich class to perform quantile regression for data in the $(0,1)$ interval (not only for median regression). The advantage of RPGJSB1$_q$ model is that $\alpha$, for a fixed quantile $\psi$, controls the shape of the distribution (different $\alpha$'s produce different shapes). However, in this parametrization the shape of the model also depends on $\psi$. As we will perform regression on $\psi$, this indicates that the shape of the quantile depend on the covariates. A second problem is the computational costs, because evaluate \ref{pdf.RPGJSB1} can be hard to compute for some combinations of $g$ and $Q$. On the other hand, the advantage of RPGJSB2$_q$ is the parsimonious (because one parameters is not estimated) and the reduction in the computational costs, because $\alpha$ is considered fixed. However, in the RPGJSB2$_q$ model the shape of the distribution is maintained (because the model belongs to the location-scale family of distributions) because such shape is ``fixed''.\\
Figure \ref{d.RPGJSB1} shows the density function for the RPGJSB1$_q(\psi,\delta=1,\alpha)$ model with logit link and $G=\Phi$ under different combinations of $q$, $\psi$ and $\alpha$. From Figure \ref{d.RPGJSB1}, note that the proposed model is very flexible since its density can assume different shapes.

\section{Inference and its associated diagnostic analysis}\label{sec:3}
\noindent

In this section, we discuss some aspects related to the inference, residuals and diagnostic analysis of the RPGJSB1$_q$ and RPGJSB2$_q$ quantile regression models.

\subsection{Inference}
\noindent

Suppose the $100\times q$th quantile $\psi$ for the RPGJSB1$_q$ model and the dispersion parameter $\delta$ satisfies the following functional relations
\begin{equation}
Q(\psi_i) = \eta_{1i} = \mathbf{x}^\top_i\bm{\beta} \quad \textrm{and} \quad \log(\delta_i) = \eta_{2i} = \mathbf{z}^\top_i\bm{\nu}, \label{cs1}
\end{equation}
or
\begin{equation}
Q(\xi_i) = \eta_{1i} = \mathbf{x}^\top_i\bm{\beta} \quad \textrm{and} \quad \log(\delta_i) = \eta_{2i} = \mathbf{z}^\top_i\bm{\nu}, \label{cs2}
\end{equation}
for the RPGJSB2$_q$ model, where $\bm{\beta} = (\beta_1, \ldots, \beta_p)^\top$ and $\bm{\nu} = (\nu_1, \ldots, \nu_r)^\top$
are vectors of unknown regression coefficients which are assumed to be functionally independent,
$\bm{\beta} \in \mathbb{R}^p$ and $\bm{\nu} \in \mathbb{R}^r$, with $p + r < n$,
$\eta_{1i}$ and $\eta_{2i}$ are the linear predictors, and $\mathbf{x}_i = (x_{i1}, \ldots, x_{ip})^\top$
and $\mathbf{z}_i = (z_{i1}, \ldots, z_{ir})^\top$ are observations on $p$ and $r$ known regressors, for $i = 1, \ldots, n$. Furthermore, we assume that the covariate matrices $\mathbf{X} = (\mathbf{x}_1, \ldots, \mathbf{x}_n)^\top$ and $\mathbf{Z} = (\mathbf{z}_1, \ldots, \mathbf{z}_n)^\top$ have rank $p$ and $r$, respectively. The log-likelihood function for the RPGJSB1$_q$
model is given by
\begin{align}
\ell_1({\bm \theta})&=\sum_{i=1}^n \bigg\{ \log(\delta_i) +\log(\alpha) +(\alpha-1)\log\left\{ G\left(\delta_i[Q(y_i)-Q(\psi_i)]+x_q^*(\alpha)\right)\right\}\nonumber \\
&~~~~~~~~~~~~\log\left\{g\left(\delta_i[Q(y_i)-Q(\psi_i)]+x_q^*(\alpha)\right)\right\}+\log \left|\frac{d Q(y_i)}{d y_i}\right|\bigg\}, \label{log.like1}
\end{align}
whereas for the RPGJSB2$_q$ is given by
\begin{align}
\ell_2({\bm \theta})&=\sum_{i=1}^n \bigg\{ \log(\delta_i) +\log(\alpha) +(\alpha-1)\log\left[G\left(\delta_i[Q(y_i)-Q(\xi_i)]\right)\right]\nonumber \\
&~~~~~~~~~~~~\log\left\{g\left(\delta_i[Q(y_i)-Q(\xi_i)]\right)\right\}+\log \left|\frac{d Q(y_i)}{d y_i}\right|\bigg\}.\label{log.like2}
\end{align}
Note that ${\bm \theta}=({\bm \beta}^\top,{\bm \nu}^\top,\alpha)$ and ${\bm \theta}=({\bm \beta}^\top,{\bm \nu}^\top)$ is the vector of parameters for the RPGJSB1$_q$ and RPGJSB2$_q$ models, respectively. The ML estimator of ${\bm \theta}$, say $\widehat{\bm \theta}$, is obtained maximizing equation (\ref{log.like1}) or (\ref{log.like2}), depending on the considered model are presented in Section . We considered the maximization procedure based on the Broyden-Fletcher-Goldfarb-Shanno (BFGS) method
initialized with a vector of zeros. To validate a solution, we checked: i) If the convergence is attached
and; ii) if the determinant of the hessian such matrix is positive. If the two conditions are not satisfied,
we rerun the procedure based initialized with a random vector generated by independent standard normal variables
until i) and ii) are satisfied. Under usual regularity conditions (see Cox and Hinkley, 1974) ${\bm \theta}$ is consistent. Moreover,
\[
{\bm \imath}^{-1}(\widehat{\bm \theta})\left[\widehat{\bm \theta}-{\bm \theta}\right]\stackrel{\mathcal{D}}{\rightarrow} N_{p+r}\left(\mathbf{0}_{p+r},\mathbf{I}_{p+r}\right), \quad \mbox{as } n\rightarrow +\infty,
\]
where ${\bm \imath}(\widehat{\bm \theta})=-\partial^2 \ell_l({\bm \theta})/\partial {\bm \theta}\partial {\bm \theta^\top}\big|_{{\bm \theta}=\widehat{\bm \theta}}$ is minus the estimated hessian matrix for the RPGJSB1$_q$ ($l=1$) and RPGJSB2$_q$ ($l=2$) models, respectively.

\subsection{Residuals}
\noindent

In order to assess if the posited model is correct, we will consider the randomized quantile residuals (RQRs) proposed by \citet{Dunn}.
For the $\mbox{RPGJSB1}_q$ model, such residuals are given by
\[
\widehat{r}_i=\Phi^{-1}\left([G(\widehat{\delta}_i [Q(y_i)-Q(\widehat{\psi}_i)]+x^*_{q}(\widehat{\alpha}))]^{\widehat{\alpha}}\right), \quad i=1,\ldots,n,
\]
whereas for the $\mbox{RPGJSB2}_q$ model, the RQRs are given by
\[
\widehat{r}_i=\Phi^{-1}\left([G(\widehat{\delta}_i [Q(y_i)-Q(\widehat{\xi}_i)])]^\alpha(q)\right), \quad i=1,\ldots,n.
\]
$\widehat{\delta}_i$, $\widehat{\xi}_i$ and $\widehat{\psi}_i$, $i=1,\ldots,n$, correspond to the expressions in equations (\ref{cs1}) and (\ref{cs2}) evaluated in $\widehat{\bm \beta}$ and $\widehat{\bm \nu}$, for each model, respectively.
If the model is correctly specified, the distribution of $\widehat{r}_1,\ldots,\widehat{r}_n$ is standard normal, which can be validated considering different normality tests, such as Kolmogorov-Smirnov (KS), Shapiro-Wilks (SW), Anderson-Darling (AD) and the Cram\'er-Von-Mises (CVM) tests. See \cite{yap11} for a discussion about such tests.

\subsection{Local influence}\label{influence}
\noindent

The local influence method suggested by \cite{Cook86} evaluates the simultaneous effect of observations on the ML estimator without removing it from the data set, based on the curvature of the plane of the log-likelihood function. Consider $\ell_1(\bm{\theta}_1;\bm{w})$ and $\ell_2(\bm{\theta}_2;\bm{w})$ the log-likelihood functions corresponding to the RPGJSB1$_q$ and RPGJSB2$_q$ models, respectively, but now perturbed by $\bm{w}$, a vector of perturbations.
$\bm{w}$ belongs to a subset $\Omega\in\mathbb{R}^n$ and $\bm{w}_0$ is a non-perturbed $n\times1$ vector, such that $\ell_l(\bm{\theta};\bm{w}_0)=\ell_l(\bm{\theta})$, for all $\bm{\theta}$, $l=1,2$. In this case, the likelihood displacement (LD) is $LD(\bm{\theta})=2(\ell_l(\widehat{\bm{\theta}})-\ell_l(\widehat{\bm{\theta}}_{\bm{w}}))$, where $\widehat{\bm{\theta}}_{\bm{w}}$ denotes the ML estimate of $\bm{\theta}$ on the perturbed regression models, that is, $\widehat{\bm{\theta}}_{\bm{w}}$ is obtained from $\ell_l(\bm{\theta};\bm{w})$. Note that $\ell_l(\bm{\theta};\bm{w})$ can be used to assess the influence of the perturbation of the ML estimate.
Cook (1986) showed that the normal curvature for $\widehat{\bm{\theta}}$ in the direction $\bm{d}$, with $||\bm{d}||=1$, is expressed as $C_{\bm{d}}(\widehat{\bm{\theta}})=2|\bm{d}^{\top}\nabla^{\top}\Sigma(\widehat{\bm{\theta}})^{-1}\nabla\bm{d}|$, where $\nabla$ is a $(p+r)\times n$ matrix of perturbations with elements $\nabla_{ji}=\partial^2\ell_l(\bm{\theta};\bm{w})/\partial\bm{\theta}_{j}\partial\bm{w}_i$, evaluated at $\bm{\theta}=\widehat{\bm{\theta}}$ and $\bm{w}=\bm{w}_0$, for $j=1,\ldots,p+r$ and $i=1,\ldots,n$. A local influence diagnostic is generally based on index plots. For example, denoting $\Sigma(\bm{\theta})$ the observed Fisher information matrix, the index graph of the eigenvector $\bm{d}_{max}$ corresponding to the maximum eigenvalue of $\bm{B}(\bm{\theta})=-\nabla^{\top}\Sigma(\bm{\theta})^{-1}\nabla$, say $C_{\bm{d}_{max}}(\bm{\theta})$, evaluated at $\bm{\theta}=\widehat{\bm{\theta}}$, can detect those cases that, under small perturbations, exert a strong influence on LD$(\bm{\theta})$.
Another important direction of interest is $\bm{d}_i=\bm{e}_{in}$, which corresponds to the direction of the case $i$, where $\bm{e}_{in}$ is an $n\times1$ vector of zeros with value equal to one at the $i$th position, that is, $\{\bm{e}_{in}, 1\leq i\leq n\}$ is the canonical basis of $\mathbb{R}^n$. In this case, the normal curvature is $C_{i}(\bm{\theta})=2|b_{ii}|$, where $b_{ii}$ is the $i$th diagonal element of $\bm{B}(\bm{\theta})$ given above, for $i=1,\ldots,n$, evaluated $\bm{\theta}=\widehat{\bm{\theta}}$.
If $C_{i}(\widehat{\bm{\theta}})>2\bar{C}(\widehat{\bm{\theta}})$, where $\bar{C}(\widehat{\bm{\theta}})=\sum_{i=1}^nC_{i}(\widehat{\bm{\theta}})/n$, it indicates case $i$ as potentially influential. This procedure is called total local influence of the case $i$ and can be carried out for $\bm{\theta}$, $\bm{\beta}$ or $\bm{\nu}$, which are denoted by $C_{i}(\bm{\theta})$, $C_{i}(\bm{\beta})$ and $C_{i}(\bm{\nu})$, respectively. We calculate the matrix $\nabla$ for three different perturbation schemes, namely: case weighting perturbation, response perturbation and explanatory variable perturbation.

\subsubsection{Perturbation of the case weights}
\noindent

In this case the perturbed log-likelihood function is given by $\ell_l(\bm{\theta};\bm{w})=\sum_{i=1}^n w_i\ell_l(\bm{\theta})$ for RPGJSB1$_q$ ($l=1$) and RPGJSB2$_q$ ($l=2$), respectively, with $0\leq w_i\leq1$, for $i=1,\ldots,n$, and $\bm{w}_0=\bm{1}^{\top}$ (all-ones vector). Hence, the perturbation matrices for the RPGJSB1$_q$ and RPGJSB2$_q$ models are given by
\begin{equation*}
\widehat{\nabla}_1=\left(
                 \begin{array}{cc}
                   \bm{X}^{\top}\widehat{\bm{D}}_1\widehat{\bm{D}}_3\\
                   \bm{Z}^{\top}\widehat{\bm{D}}_2\widehat{\bm{D}}_4\\
                 \end{array}
               \right)\quad \quad \mbox{and} \quad \quad \widehat{\nabla}_2=\left(
                 \begin{array}{cc}
                   \bm{X}^{\top}\widehat{\bm{D}}_5\widehat{\bm{D}}_7\widehat{\bm{D}}_9\\
                   \bm{Z}^{\top}\widehat{\bm{D}}_6\widehat{\bm{D}}_8\widehat{\bm{D}}_9\\
                 \end{array}
               \right),
\end{equation*}
respectively, with $\bm{D}_1=[a_i\iota_{ij}]$, $\bm{D}_2=[b_i\iota_{ij}]$, $\bm{D}_3=[\dot{d}_{\psi}\iota_{ij}]$ and $\bm{D}_4=[\dot{d}_{\delta}\iota_{ij}]$ where $a_i=\partial\psi_i/\partial\eta_{i1}$ and $b_i=\partial\delta_i/\partial\eta_{i2}$ defined from (\ref{cs1});  $\dot{d}_{\psi}=\partial\ell_1(\psi_i,\delta_i)/\partial\psi_i$, $\dot{d}_{\delta}=\partial\ell_1(\psi_i,\delta_i)/\partial\delta_i$ defined from the RPGJSB1$_q$ model and $\iota_{ij}$ is the Kronecker delta for $i,j=1,2,\ldots,n$.
Similarly, $\bm{D}_5=[c_i\iota_{ij}]$, $\bm{D}_6=[d_i\iota_{ij}]$, $\bm{D}_7=[\dot{d}_{\xi}\iota_{ij}]$, $\bm{D}_8=[\dot{d}_{\delta}\iota_{ij}]$ and $\bm{D}_9=[\dot{d}_{\alpha}\iota_{ij}]$ where $c_i=\partial\xi_i/\partial\eta_{i1}$ and $d_i=\partial\delta_i/\partial\eta_{i2}$ defined from (\ref{cs2}); $\dot{d}_{\xi}=\partial\ell_2(\xi_i,\delta_i,\alpha)/\partial\xi_i$, $\dot{d}_{\delta}=\partial\ell_2(\xi_i,\delta_i,\alpha)/\partial\delta_i$ and $\dot{d}_{\alpha}=\partial\ell_2(\xi_i,\delta_i,\alpha)/\partial\alpha$ defined from the RPGJSB2$_q$ model.

\subsubsection{Perturbation of the response}
\noindent

Now consider an multiplicative perturbation of the $i$th response by making $y_i(w_i)=y_iw_is_y$, where $s_y$ represents a scale factor
and $w_i\in\mathbb{R}$, for $i=1,\ldots,n$. Then, under the scheme of response perturbation, the log-likelihood function is given by $\ell_1({\bm \theta};\bm{w})=\sum_{i=1}^n\ell_1(\psi_i,\delta_i,\alpha;\bm{w})$ for the RPGJSB1$_q$ model and $\ell_2({\bm\theta};\bm{w})=\sum_{i=1}^n\ell_2(\xi_i,\delta_i;\bm{w})$ for the RPGJSB2$_q$ model, where
\begin{eqnarray*}
\ell_1(\psi_i,\delta_i,\alpha;\bm{w})&=&(\alpha-1)\log(G(\tau_{1i}))+\log(\alpha\delta_i)+\log(g(\tau_{1i}))+\log(|w_is_y\dot{Q}_{y}(y_iw_is_y)|)\nonumber\\
\ell_2(\xi_i,\delta_i;\bm{w})&=&(\alpha-1)\log(G(\tau_{2i}))+\log(\alpha\delta_i)+\log(g(\tau_{2i}))+\log(|w_is_y\dot{Q}_{y}(y_iw_is_y)|)
\end{eqnarray*}
with $\tau_{1i}=\delta_i(Q(y_iw_is_y)-Q(\psi_i))$ and $\tau_{2i}=\delta_i(Q(y_iw_is_y)-Q(\xi_i))+x^*_{q}(\alpha)$.

The disturbance matrices of the RPGJSB1$_q$ and RPGJSB2$_q$ models here take the form
\begin{equation*}
\widehat{\nabla}_1=\left(
                 \begin{array}{cc}
                   \bm{X}^{\top}\widehat{\bm{D}}_1\widehat{\bm{D}}_{10}\bm{S}\\
                   \bm{Z}^{\top}\widehat{\bm{D}}_2\widehat{\bm{D}}_{11}\bm{S}\\
                 \end{array}
               \right)\quad \quad \mbox{and} \quad \quad \widehat{\nabla}_2=\left(
                 \begin{array}{cc}
                   \bm{X}^{\top}\widehat{\bm{D}}_5\widehat{\bm{D}}_{12}\widehat{\bm{D}}_{14}\bm{S}\\
                   \bm{Z}^{\top}\widehat{\bm{D}}_6\widehat{\bm{D}}_{13}\widehat{\bm{D}}_{14}\bm{S}\\
                 \end{array}
               \right)
\end{equation*}
where $\bm{S}= [s_y\iota_{ij}]$, the $i$th element of matrices $\bm{D}_{10}$ and $\bm{D}_{11}$ for model RPGJSB1$_q$ and matrices $\bm{D}_{12}$, $\bm{D}_{13}$ and $\bm{D}_{14}$ for model RPGJSB2$_q$ are detailed in Section A.1 of the supplementary material.

\subsubsection{Perturbation of the predictor}
\noindent

Now consider an multiplicative perturbation of the $i$th predictor by making $x_{i}(w_i)=\bm{x}^{\top}_iw_i$ and $z_{i}(w_i)=\bm{z}^{\top}_iw_i$, for $w_i\in\mathbb{R}$, $i=1,\ldots,n$. Then, under the scheme of prediction perturbation, the log-likelihood function is given by $\ell_1({\bm \theta};\bm{w})=\sum_{i=1}^n\ell_1(\psi_i^\star,\delta_i^\star)$ for the RPGJSB1$_q$ model and $\ell_2({\bm \theta};\bm{w})=\sum_{i=1}^n\ell_2(\xi_i^\star,\delta_i^\star,\alpha)$ for the RPGJSB2$_q$ model, where $Q(\psi_i^\star)=\bm{x}^{\top}_i\bm{\beta}w_i$ and $\delta_i^\star=\exp\{\bm{z}^{\top}_i\bm{\nu}w_i\}$ for the RPGJSB1$_q$ model and
$Q(\xi_i^\star)=\bm{x}^{\top}_i\bm{\beta}w_i$ and $\delta_i^\star=\exp\{\bm{z}^{\top}_i\bm{\nu}w_i\}$ for the RPGJSB2$_q$ model.

The disturbance matrices of RPGJSB1$_q$ and RPGJSB2$_q$ models here take the form
\begin{equation*}
\widehat{\nabla}_1=\left(
                 \begin{array}{cc}
                   \bm{X}^{\top}\widehat{\bm{D}}_{15}\\
                   \bm{Z}^{\top}\widehat{\bm{D}}_{16}\\
                 \end{array}
               \right)\quad \quad \mbox{and} \quad \quad \widehat{\nabla}_2=\left(
                 \begin{array}{cc}
                   \bm{X}^{\top}\widehat{\bm{D}}_{17}\widehat{\bm{D}}_{19}\\
                   \bm{Z}^{\top}\widehat{\bm{D}}_{18}\widehat{\bm{D}}_{19}\\
                 \end{array}
               \right)
\end{equation*}
where the $i$th elements of matrices $\bm{D}_{15}$ and $\bm{D}_{16}$ for RPGJSB1$_q$ model and matrices $\bm{D}_{17}$, $\bm{D}_{18}$ and $\bm{D}_{19}$ for RPGJSB2$_q$ model are detailed in Section A.2. of the supplementary material.

\section{Simulation studies}\label{sec:4}
\noindent

In this section, we present a simulation study to assess the performance of ${\bm\theta} = ({\bm \beta},{\bm \nu},\alpha)^\top$ under different scenarios. First, we assume that $G$ and the link function are correctly specified.
The data were drawn motivated by the scheme for the anchoveta data set presented in Section C of the supplementary material. We considered $\mathbf{x}_i=\mathbf{z}_i$, where both matrices includes an intercept and a covariate. Such covariates were drawn from the $U(-5.478, -2.305)$ distribution. We considered the logistic and normal models for $G$ and the logit and loglog link functions. The true values for parameters were considered as the estimated parameters for three values for $q=\{0.1, 0.5, 0.9\}$. We also considered three sample sizes: $100, 200$ and $500$.

\begin{table}[!htbp]
\caption{True parameters used for simulation studies.}
\label{true.values}
\begin{center}
\begin{tabular}{cccccrrcccrr}
\hline
           &            &                                  \multicolumn{ 5}{c}{logistic} &                                    \multicolumn{ 5}{c}{normal} \\

       link &          $q$ &      $\beta_0$ &      $\beta_1$ &        $\nu_0$ &        $\nu_1$ &  $\log(\alpha)$ &      $\beta_0$ &      $\beta_1$ &        $\nu_0$ &        $\nu_1$ &  $\log(\alpha)$ \\
\hline

      logit &        0.1 &        4.9 &        2.6 &        2.2 &        0.4 &       $-$0.7 &        4.4 &        2.4 &        1.5 &        0.3 &       $-$1.4 \\

            &        0.5 &        4.8 &        2.1 &        2.2 &        0.4 &       $-$0.7 &        4.6 &        2.1 &        1.5 &        0.3 &       $-$1.4 \\

            &        0.9 &        4.7 &        1.8 &        2.2 &        0.4 &       $-$0.7 &        4.8 &        1.9 &        1.5 &        0.3 &       $-$1.4 \\
\hline

     loglog &        0.1 &        1.3 &        0.8 &        0.8 &       $-$0.3 &        0.1 &        1.2 &        0.7 &       $-$0.1 &       $-$0.3 &        1.1 \\

            &        0.5 &        2.1 &        0.9 &        1.0 &       $-$0.2 &        0.1 &        2.0 &        0.9 &        0.0 &       $-$0.3 &        1.0 \\

            &        0.9 &        2.8 &        1.0 &        1.1 &       $-$0.2 &        0.1 &        2.8 &        1.0 &        0.1 &       $-$0.2 &        1.0 \\
\hline

\end{tabular}
\end{center}
\end{table}

As mentioned previously, to validate a solution, we checked: If the convergence is attached and if the determinant of the hessian such matrix is positive. If the two conditions are not satisfied, we rerun the procedure initialized with a random vector generated by independent standard normal variables until both conditions are satisfied.
For each combination of $G$, link, $q$ and sample size, we considered 5,000 replicates and in each case the estimation is performed based on the same $G$ and link function. Based on the 10,000 replicates, we report the bias for each estimator, the standard error of the estimates ($SE_1$), the mean of the estimated standard errors ($SE_2$) and the 95\% coverage probabilities (CP).
Tables \ref{estimation} and \ref{estimation2} summarizes such results. Note that the bias of the parameters is reduced and the terms $SE_1$ and $SE_2$ are closer when $n$ is increased, suggesting that the estimators are consistent in finite samples. Additionally, when the sample size is increased the CP are closer to the nominal value used. Finally, Table \ref{estimation3} presents the percentage of times where the algorithm converges when is initialized with a vector of zeros. Note that the maximization procedure converged at least in 89.43\% of the generated samples and such percentages are increased when the sample size is increased.

\begin{table}[!h]
\caption{Recovery parameters when $G$ and the link are correctly specified (case $G$ is the cdf of the logistic distribution).}
\label{estimation}
\begin{center}
\resizebox{\linewidth}{!}{
\begin{tabular}{ccccrcccrcccrccc}
\hline
         $G$ &       link &          $q$ &  parameter &       bias &        $SE_1$ &        $SE_2$ &         CP &       bias &        $SE_1$ &        $SE_2$ &         CP &       bias &        $SE_1$ &        $SE_2$ &         CP \\
\hline
  logistic &      logit &        0.1 &      $\beta_0$ &     $-$0.034 &      0.753 &      0.728 &      0.938 &     $-$0.017 &      0.538 &      0.529 &      0.946 &     $-$0.007 &      0.345 &      0.339 &      0.946 \\

           &            &            &      $\beta_1$ &     $-$0.015 &      0.238 &      0.229 &      0.934 &     $-$0.007 &      0.166 &      0.163 &      0.942 &     $-$0.003 &      0.104 &      0.102 &      0.947 \\

           &            &            &        $\nu_0$ &      0.041 &      0.381 &      0.367 &      0.935 &      0.020 &      0.269 &      0.263 &      0.942 &      0.009 &      0.171 &      0.170 &      0.947 \\

           &            &            &        $\nu_1$ &     $-$0.001 &      0.088 &      0.085 &      0.939 &      0.000 &      0.061 &      0.060 &      0.946 &      0.000 &      0.039 &      0.038 &      0.946 \\

           &            &            &   $\log(\alpha)$ &     $-$0.004 &      0.355 &      0.331 &      0.947 &     $-$0.002 &      0.232 &      0.224 &      0.946 &     $-$0.002 &      0.140 &      0.138 &      0.948 \\
\cline{3-16}
           &            &        0.5 &      $\beta_0$ &     $-$0.017 &      0.485 &      0.472 &      0.941 &      0.001 &      0.322 &      0.319 &      0.946 &    $-$0.003 &      0.204 &      0.205 &      0.950 \\

           &            &            &      $\beta_1$ &     $-$0.005 &      0.146 &      0.142 &      0.939 &      0.000 &      0.096 &      0.095 &      0.946 &     $-$0.001 &      0.061 &      0.061 &      0.949 \\

           &            &            &        $\nu_0$ &      0.046 &      0.452 &      0.443 &      0.946 &      0.027 &      0.296 &      0.294 &      0.948 &      0.007 &      0.183 &      0.182 &      0.949 \\

           &            &            &        $\nu_1$ &      0.002 &      0.107 &      0.106 &      0.946 &      0.002 &      0.068 &      0.068 &      0.949 &      0.000 &      0.042 &      0.042 &      0.951 \\

           &            &            &   $\log(\alpha)$ &      0.004 &      0.352 &      0.331 &      0.952 &     $-$0.001 &      0.231 &      0.224 &      0.948 &      0.000 &      0.142 &      0.139 &      0.947 \\
\cline{3-16}

           &            &        0.9 &      $\beta_0$ &     $-$0.001 &      0.620 &      0.591 &      0.930 &     $-$0.002 &      0.369 &      0.363 &      0.942 &      0.002 &      0.237 &      0.236 &      0.950 \\

           &            &            &      $\beta_1$ &      0.004 &      0.177 &      0.169 &      0.932 &      0.002 &      0.112 &      0.111 &      0.943 &      0.001 &      0.072 &      0.072 &      0.948 \\

           &            &            &        $\nu_0$ &      0.060 &      0.461 &      0.443 &      0.943 &      0.024 &      0.289 &      0.283 &      0.943 &      0.007 &      0.184 &      0.182 &      0.946 \\

           &            &            &        $\nu_1$ &      0.006 &      0.103 &      0.100 &      0.938 &      0.002 &      0.066 &      0.065 &      0.946 &      0.000 &      0.043 &      0.043 &      0.945 \\

           &            &            &   $\log(\alpha)$ &      0.010 &      0.362 &      0.334 &      0.946 &      0.003 &      0.234 &      0.224 &      0.949 &      0.002 &      0.140 &      0.139 &      0.949 \\
\cline{2-16}

           &     loglog &        0.1 &      $\beta_0$ &      0.008 &      0.175 &      0.168 &      0.931 &      0.002 &      0.116 &      0.113 &      0.938 &      0.000 &      0.071 &      0.071 &      0.949 \\

           &            &            &      $\beta_1$ &      0.001 &      0.039 &      0.037 &      0.935 &      0.000 &      0.026 &      0.025 &      0.937 &      0.000 &      0.016 &      0.016 &      0.948 \\

           &            &            &        $\nu_0$ &      0.020 &      0.413 &      0.398 &      0.944 &      0.005 &      0.280 &      0.275 &      0.946 &     $-$0.001 &      0.165 &      0.167 &      0.950 \\

           &            &            &        $\nu_1$ &      0.000 &      0.096 &      0.092 &      0.939 &     $-$0.002 &      0.067 &      0.065 &      0.942 &     $-$0.001 &      0.039 &      0.039 &      0.949 \\

           &            &            &   $\log(\alpha)$ &      0.153 &      1.175 &      2.515 &      0.964 &      0.035 &      0.349 &      0.324 &      0.961 &      0.014 &      0.178 &      0.174 &      0.956 \\
\cline{3-16}

           &            &        0.5 &      $\beta_0$ &     $-$0.002 &      0.130 &      0.128 &      0.944 &     $-$0.003 &      0.090 &      0.090 &      0.951 &      0.001 &      0.061 &      0.061 &      0.946 \\

           &            &            &      $\beta_1$ &      0.000 &      0.031 &      0.030 &      0.945 &     $-$0.001 &      0.021 &      0.021 &      0.949 &      0.000 &      0.014 &      0.014 &      0.947 \\

           &            &            &        $\nu_0$ &      0.007 &      0.386 &      0.376 &      0.944 &      0.003 &      0.264 &      0.261 &      0.947 &      0.006 &      0.175 &      0.175 &      0.950 \\

           &            &            &        $\nu_1$ &     $-$0.003 &      0.093 &      0.091 &      0.945 &     $-$0.002 &      0.063 &      0.062 &      0.949 &      0.000 &      0.041 &      0.041 &      0.950 \\

           &            &            &   $\log(\alpha)$ &      0.143 &      1.070 &      2.042 &      0.965 &      0.041 &      0.306 &      0.290 &      0.962 &      0.012 &      0.177 &      0.174 &      0.951 \\
\cline{3-16}

           &            &        0.9 &      $\beta_0$ &     $-$0.005 &      0.178 &      0.175 &      0.939 &     $-$0.004 &      0.141 &      0.139 &      0.942 &     $-$0.002 &      0.082 &      0.082 &      0.947 \\

           &            &            &      $\beta_1$ &     $-$0.001 &      0.042 &      0.041 &      0.940 &     $-$0.001 &      0.033 &      0.032 &      0.943 &      0.000 &      0.019 &      0.019 &      0.947 \\

           &            &            &        $\nu_0$ &      0.012 &      0.387 &      0.374 &      0.940 &      0.010 &      0.296 &      0.288 &      0.944 &      0.004 &      0.174 &      0.173 &      0.949 \\

           &            &            &        $\nu_1$ &     $-$0.002 &      0.094 &      0.091 &      0.940 &      0.000 &      0.071 &      0.069 &      0.944 &      0.000 &      0.041 &      0.041 &      0.949 \\

           &            &            &   $\log(\alpha)$ &      0.133 &      0.968 &      1.596 &      0.965 &      0.042 &      0.311 &      0.290 &      0.961 &      0.014 &      0.177 &      0.174 &      0.952 \\
\hline
\end{tabular}
}
\end{center}
\end{table}

\begin{table}[!h]
\caption{Recovery parameters when $G$ and the link are correctly specified (case $G$ is the cdf of the normal distribution).}
\label{estimation2}
\begin{center}
\resizebox{\linewidth}{!}{
\begin{tabular}{ccccrcccrcccrccc}
\hline
         $G$ &       link &          $q$ &  parameter &       bias &        $SE_1$ &        $SE_2$ &         CP &       bias &        $SE_1$ &        $SE_2$ &         CP &       bias &        $SE_1$ &        $SE_2$ &         CP \\
\hline
    normal &      logit &        0.1 &      $\beta_0$ &     $-$0.004 &      0.725 &      0.711 &      0.939 &      0.000 &      0.470 &      0.473 &      0.952 &     $-$0.002 &      0.289 &      0.291 &      0.951 \\

           &            &            &      $\beta_1$ &     $-$0.005 &      0.202 &      0.198 &      0.939 &     $-$0.002 &      0.133 &      0.134 &      0.951 &     $-$0.002 &      0.081 &      0.082 &      0.949 \\

           &            &            &        $\nu_0$ &      0.912 &      2.171 &      0.730 &      0.847 &      0.243 &      1.024 &      0.450 &      0.946 &      0.045 &      0.270 &      0.248 &      0.954 \\

           &            &            &        $\nu_1$ &      0.000 &      0.083 &      0.079 &      0.932 &      0.001 &      0.055 &      0.054 &      0.945 &      0.000 &      0.032 &      0.032 &      0.951 \\

           &            &            &   $\log(\alpha)$ &     $-$1.763 &      4.569 &      1.650 &      0.867 &     $-$0.462 &      2.167 &      1.019 &      0.960 &     $-$0.082 &      0.612 &      0.565 &      0.956 \\
\cline{3-16}

           &            &        0.5 &      $\beta_0$ &     $-$0.006 &      0.464 &      0.452 &      0.942 &     $-$0.005 &      0.330 &      0.324 &      0.945 &      0.002 &      0.196 &      0.194 &      0.946 \\

           &            &            &      $\beta_1$ &     $-$0.004 &      0.135 &      0.131 &      0.940 &     $-$0.002 &      0.094 &      0.092 &      0.941 &      0.000 &      0.056 &      0.056 &      0.946 \\

           &            &            &        $\nu_0$ &      0.944 &      2.251 &      0.703 &      0.841 &      0.215 &      0.966 &      0.450 &      0.949 &      0.040 &      0.281 &      0.250 &      0.952 \\

           &            &            &        $\nu_1$ &      0.002 &      0.082 &      0.079 &      0.939 &      0.001 &      0.056 &      0.055 &      0.947 &      0.000 &      0.034 &      0.033 &      0.950 \\

           &            &            &   $\log(\alpha)$ &     $-$1.806 &      4.729 &      1.597 &      0.862 &     $-$0.398 &      2.046 &      1.012 &      0.961 &     $-$0.071 &      0.625 &      0.564 &      0.954 \\
\cline{3-16}

           &            &        0.9 &      $\beta_0$ &     $-$0.028 &      0.595 &      0.550 &      0.910 &     $-$0.001 &      0.393 &      0.375 &      0.934 &     $-$0.004 &      0.244 &      0.242 &      0.947 \\

           &            &            &      $\beta_1$ &     $-$0.002 &      0.165 &      0.153 &      0.912 &      0.003 &      0.111 &      0.106 &      0.933 &      0.000 &      0.069 &      0.069 &      0.949 \\

           &            &            &        $\nu_0$ &      0.923 &      2.248 &      0.712 &      0.852 &      0.235 &      1.009 &      0.450 &      0.947 &      0.047 &      0.279 &      0.253 &      0.949 \\

           &            &            &        $\nu_1$ &      0.006 &      0.088 &      0.084 &      0.937 &      0.001 &      0.057 &      0.055 &      0.941 &      0.001 &      0.035 &      0.035 &      0.949 \\

           &            &            &   $\log(\alpha)$ &     $-$1.733 &      4.706 &      1.576 &      0.871 &     $-$0.434 &      2.133 &      1.008 &      0.961 &     $-$0.083 &      0.614 &      0.563 &      0.956 \\
\cline{2-16}

           &     loglog &        0.1 &      $\beta_0$ &      0.005 &      0.156 &      0.152 &      0.935 &      0.005 &      0.115 &      0.114 &      0.942 &      0.001 &      0.070 &      0.069 &      0.946 \\

           &            &            &      $\beta_1$ &      0.000 &      0.035 &      0.034 &      0.936 &      0.001 &      0.026 &      0.026 &      0.945 &      0.000 &      0.016 &      0.015 &      0.946 \\

           &            &            &        $\nu_0$ &      0.085 &      0.834 &      0.530 &      0.963 &      0.024 &      0.371 &      0.351 &      0.951 &      0.006 &      0.209 &      0.209 &      0.952 \\

           &            &            &        $\nu_1$ &     $-$0.004 &      0.077 &      0.076 &      0.942 &     $-$0.001 &      0.059 &      0.058 &      0.946 &     $-$0.001 &      0.035 &      0.035 &      0.951 \\

           &            &            &   $\log(\alpha)$ &      1.090 &     23.978 &      3.284 &      0.965 &      0.103 &      1.677 &      1.200 &      0.958 &      0.027 &      0.661 &      0.658 &      0.961 \\
\cline{3-16}

           &            &        0.5 &      $\beta_0$ &      0.002 &      0.116 &      0.114 &      0.942 &      0.000 &      0.084 &      0.083 &      0.946 &      0.000 &      0.054 &      0.053 &      0.948 \\

           &            &            &      $\beta_1$ &      0.000 &      0.026 &      0.025 &      0.942 &      0.000 &      0.019 &      0.019 &      0.947 &      0.000 &      0.012 &      0.012 &      0.947 \\

           &            &            &        $\nu_0$ &      0.123 &      0.990 &      0.539 &      0.954 &      0.017 &      0.379 &      0.348 &      0.955 &      0.009 &      0.212 &      0.212 &      0.952 \\

           &            &            &        $\nu_1$ &     $-$0.004 &      0.082 &      0.079 &      0.939 &     $-$0.002 &      0.059 &      0.059 &      0.946 &     $-$0.001 &      0.036 &      0.036 &      0.950 \\

           &            &            &   $\log(\alpha)$ &      0.612 &     16.917 &      3.114 &      0.964 &      0.091 &      1.453 &      1.150 &      0.963 &      0.017 &      0.654 &      0.645 &      0.957 \\
\cline{3-16}

           &            &        0.9 &      $\beta_0$ &     $-$0.016 &      0.224 &      0.219 &      0.935 &     $-$0.007 &      0.169 &      0.167 &      0.939 &     $-$0.004 &      0.095 &      0.095 &      0.943 \\

           &            &            &      $\beta_1$ &     $-$0.002 &      0.051 &      0.050 &      0.937 &     $-$0.001 &      0.040 &      0.039 &      0.940 &     $-$0.001 &      0.022 &      0.022 &      0.946 \\

           &            &            &        $\nu_0$ &      0.125 &      0.934 &      0.538 &      0.958 &      0.029 &      0.386 &      0.357 &      0.951 &      0.008 &      0.208 &      0.207 &      0.951 \\

           &            &            &        $\nu_1$ &      0.000 &      0.083 &      0.080 &      0.940 &      0.000 &      0.061 &      0.060 &      0.948 &      0.000 &      0.035 &      0.034 &      0.952 \\

           &            &            &   $\log(\alpha)$ &      0.428 &     12.877 &      2.696 &      0.964 &      0.088 &      1.443 &      1.192 &      0.957 &      0.034 &      0.657 &      0.647 &      0.961 \\
\hline
\end{tabular}
}
\end{center}
\end{table}

\begin{table}[!h]
\caption{Percentage of time where the maximization algorithm converges with initial value as the vector zero.}
\label{estimation3}
\begin{center}
\resizebox{\linewidth}{!}{
\begin{tabular}{ccrrrrrrrrr}
\hline
           &            &           \multicolumn{ 3}{c}{$q=0.1$} &           \multicolumn{ 3}{c}{$q=0.5$} &           \multicolumn{ 3}{c}{$q=0.9$} \\

         G &       link &     100 &     200 &     500 &     100 &     200 &     500 &     100 &     200 &     500 \\
\hline

  logistic &      logit &     100.00 &     100.00 &     100.00 &     100.00 &     100.00 &     100.00 &     100.00 &     100.00 &     100.00 \\

           &     loglog &      99.71 &     100.00 &     100.00 &      99.83 &     100.00 &     100.00 &      99.85 &     100.00 &     100.00 \\
\hline

    normal &      logit &      90.77 &      98.40 &     100.00 &      89.43 &      98.65 &      99.99 &      90.38 &      98.59 &      99.99 \\

           &     loglog &      99.43 &      99.99 &     100.00 &      99.01 &      99.98 &     100.00 &      99.05 &      99.98 &     100.00 \\
\hline

\end{tabular}
}
\end{center}
\end{table}

\newpage

\section{Data analysis}\label{sec:5}
\noindent

In this section, we present a real data set application related to the mortality rate of the COVID-19 in different countries to illustrate the performance of the RPGJSB1$_q$ and RPGJSB2$_q$ regression models. An additional application related to
the reproductive activity of the anchoveta in Chile is presented in Section C of the supplementary material.

\subsection{COVID-19 data set}
\noindent

\begin{figure}[!h]
\psfrag{-1}[c][c]{\tiny{-1}}
\psfrag{-2}[c][c]{\tiny{-2}}
\psfrag{-3}[c][c]{\tiny{-3}}
\psfrag{-4}[c][c]{\tiny{-4}}
\psfrag{-5}[c][c]{\tiny{-5}}
\psfrag{-6}[c][c]{\tiny{-6}}
\psfrag{-0.5}[c][c]{\tiny{-0.5}}
\psfrag{-1.0}[c][c]{\tiny{-1.0}}
\psfrag{-1.5}[c][c]{\tiny{-1.5}}
\psfrag{-2.0}[c][c]{\tiny{-2.0}}
\psfrag{-2.5}[c][c]{\tiny{-2.5}}
\psfrag{-3.0}[c][c]{\tiny{-3.0}}
\psfrag{8}[c][c]{\tiny{8}}
\psfrag{10}[c][c]{\tiny{10}}
\psfrag{12}[c][c]{\tiny{12}}
\psfrag{14}[c][c]{\tiny{14}}
\psfrag{16}[c][c]{\tiny{16}}
\psfrag{18}[c][c]{\tiny{18}}
\psfrag{0.66}[c][c]{\tiny{}}
\psfrag{0.64}[c][c]{\tiny{0.64}}
\psfrag{BxB}[c][c]{\scriptsize{$\log$(surface)}}
\begin{minipage}[b]{0.23\linewidth}
\psfrag{AxA}[c][c]{\scriptsize{logit(\texttt{mort})}}
\centering
\includegraphics[width=4cm]{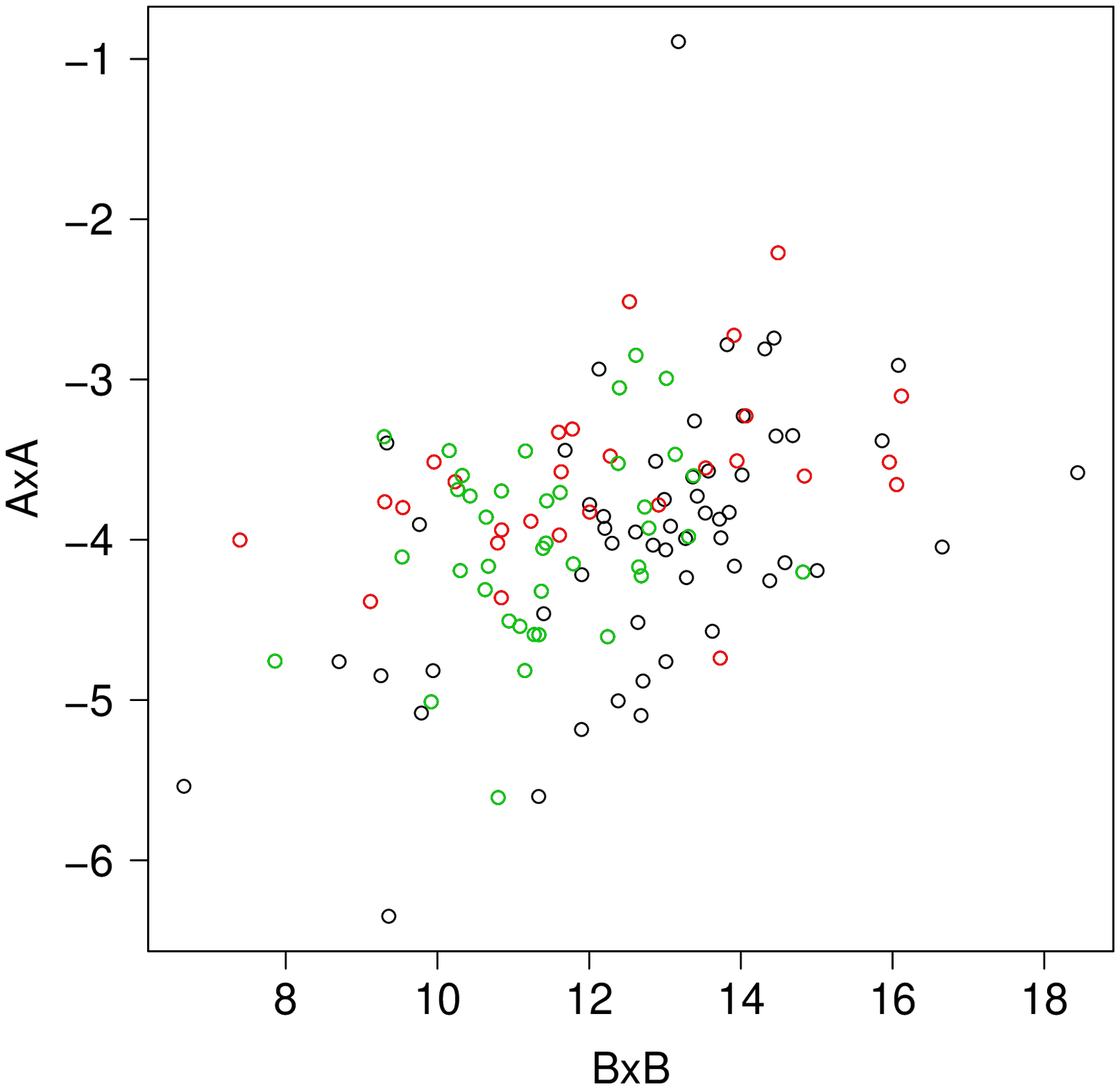}
\end{minipage} %\hfill
\hspace{0.2cm}
\begin{minipage}[b]{0.23\linewidth}
\psfrag{0.58}[c][c]{\tiny{}}
\psfrag{0.56}[c][c]{\tiny{}}
\psfrag{0.54}[c][c]{\tiny{}}
\psfrag{0.52}[c][c]{\tiny{0.52}}
\psfrag{AxA}[c][c]{\scriptsize{probit(\texttt{mort})}}
\centering
\includegraphics[width=4cm]{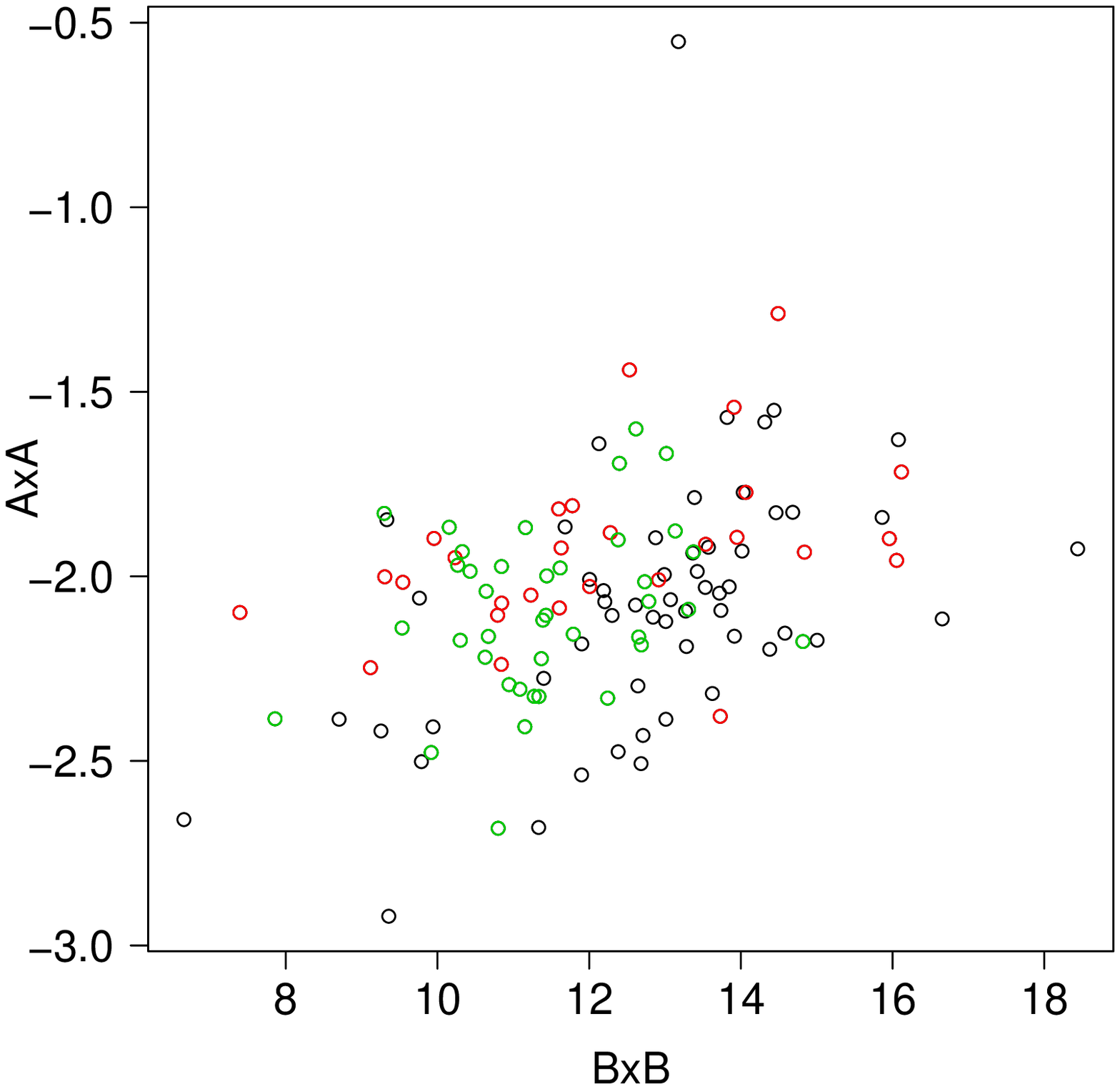}
\end{minipage}
\hspace{0.2cm}
\begin{minipage}[b]{0.23\linewidth}
\psfrag{AxA}[c][c]{\scriptsize{loglog(\texttt{mort})}}
\centering
\includegraphics[width=4cm]{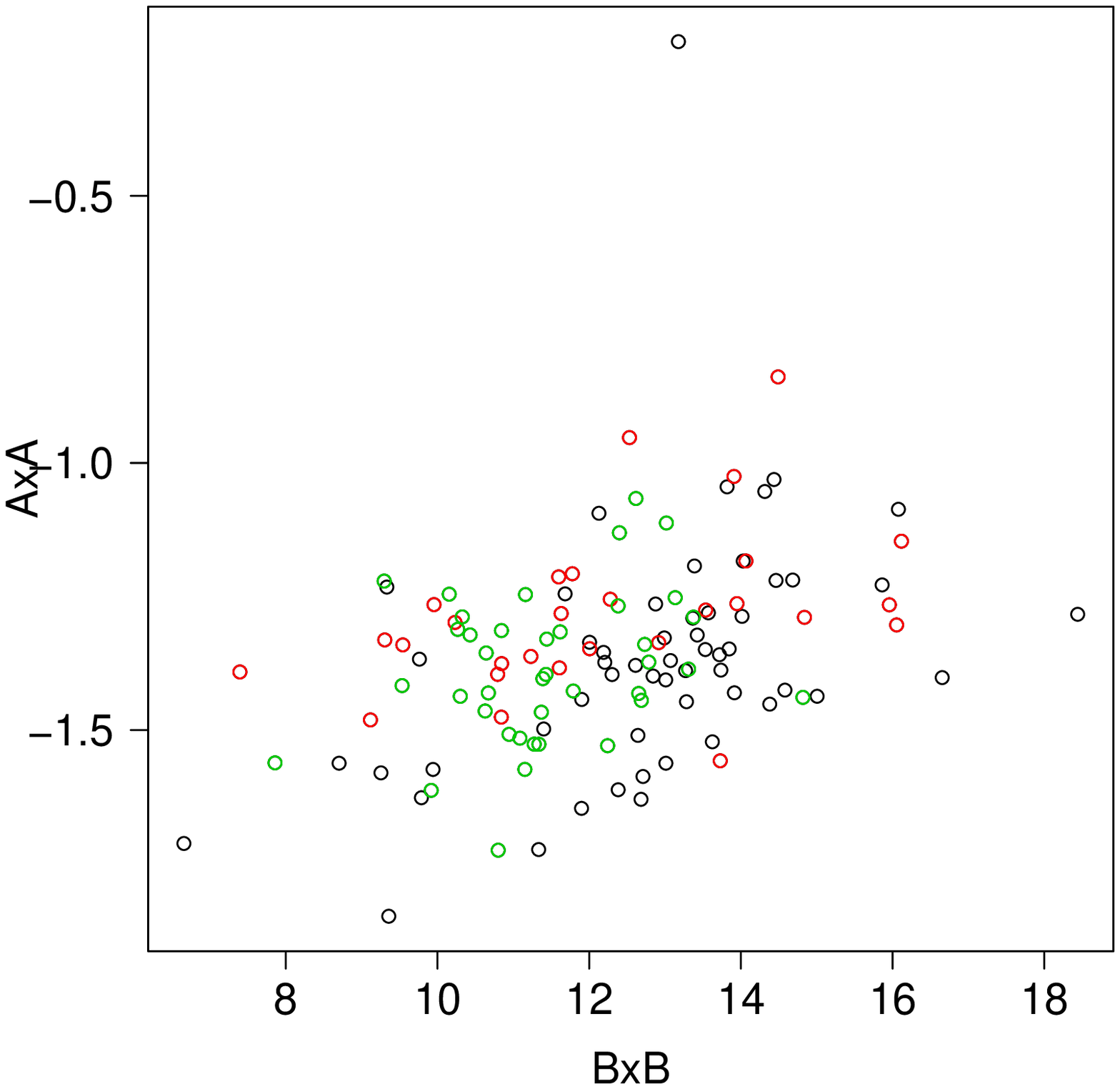}
\end{minipage} %\hfill
\hspace{0.2cm}
\begin{minipage}[b]{0.23\linewidth}
\psfrag{AxA}[c][c]{\scriptsize{cloglog(\texttt{mort})}}
\centering
\includegraphics[width=4cm]{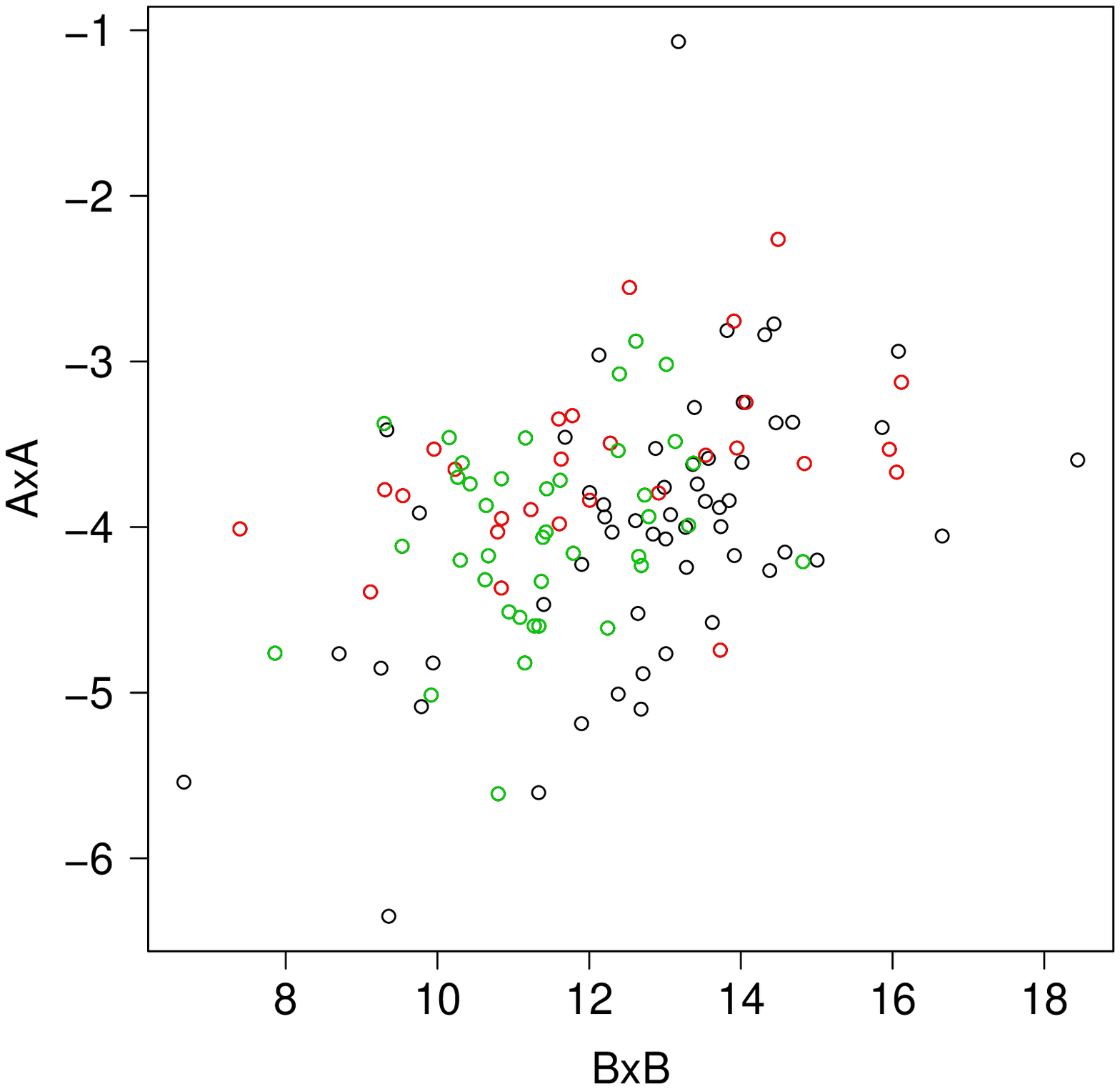}
\end{minipage} %\hfill
\caption{Descriptive plots for $Q$(\texttt{mort}) versus log(\texttt{surface}) for different link functions: logit, probit, loglog and cloglog and separated by continent: Africa, Asia u Oceania (black), America (red) and Europe (green).}\label{desc.COVID}
\end{figure}

The COVID-19 pandemic has unprecedentedly affected the entire worldall. Specifically, has yielded high mortality rates since its emergence in December 2019, generating a disequilibrium societal, economic, cultural and political. It has been shown by early studies that statistical analysis can be applied to COVID-19 problems to build predictive models that can assess risk factors and mortality rates \citep{Ji, Xi, Du}. Also the overall mortality rate has been about 5\%, while the statistics showed a rate of around 20\% for senior patients \citep{Livingston}. We consider the following information for the countries with at least 1,000 reported cases of COVID-19 and at least 100 deaths attributed to COVID-19, totalizing 123 countries at November 3, 2020.
\begin{itemize}
\item \texttt{mort}: mortality rate (reported death/reported cases). Mean=0.025, Median=0.020, standard deviation=0.028, minimum=0.002 and maximum=0.291.
\item \texttt{surface}: surface of the country (in km$^2$).
\item \texttt{population}: official estimated population of the country.
\item \texttt{cont}: continent to which the country belongs (categorized as 1: Africa, Asia u Oceania; 2: America; 3: Europe; with 56, 28 and 39 countries, respectively).
\end{itemize}
The information was taken from the World Heatlh Organization \citep{Who}.
It is of interest to model the mortality rate in terms of the surface and the continent of each country (previous analysis suggest that the population is not significative to model the mortality rate). Figure \ref{desc.COVID} shows the plots for $Q(\texttt{mort})$ for different link functions versus the log(\texttt{surface}) and separated by \texttt{cont}.

\subsubsection{Estimation}
\noindent

In view of the above, we consider to model the mortality rate using $\texttt{mort}_i\sim RPGJSB1_q(\psi_i,\delta_i,\alpha)$, with
\begin{align*}
Q(\psi_i)&=\beta_{0}+\beta_1\times \log(\mbox{surface}_i)+\beta_2 \times \texttt{America}_i+\beta_3\times \texttt{Europe}_i \qquad \mbox{and} \\
\log(\delta_i)&=\nu_{0}+\nu_1 \times \texttt{America}_i+\nu_2 \times \texttt{Europe}_i,
\end{align*}
or alternatively, $\texttt{mort}_i\sim RPGJSB2_q(\xi_i,\delta_i)$, where $Q(\xi_i)=\beta_{0}+\beta_1\times \log(\mbox{surface}_i)+\beta_2 \times \texttt{America}_i+\beta_3\times \texttt{Europe}_i$ and $\delta_i$ is modelled in the same way. In Section B.1 of the supplementary material, we present the AIC and BIC for $q$ ranging in the set $\{0.05, 0.10, \ldots, 0.90, 0.95\}$ and the RPGJSB1$_q$ and RPGJSB2$_q$ models. Note that the RPGJSB1$_q$ provides the lower AIC than the RPGJSB2$_q$ for all the considered $q$. Then, hereinafter we focused in the RPGJSB1$_q$ model, specifically where $G$ is the cdf of the logistic model and the cloglog link (which provide the lower AIC for all $q$). Table \ref{estimation.covid} and Section B.2
of the supplementary material present the estimated parameter for such model for five selected quantiles. Also are presented the KS, SW, AD and CVM tests to check the normality of the RQRs. Note that the log(\texttt{surface}) is significative to model the quantile (with a nominal level of 5\%) for all the considered $q$. This can be explained because countries with larger areas may have greater difficulties in providing medical coverage to their inhabitants in relation to countries with smaller areas.
Also the parameter related to \texttt{America} is significant in both, quantile and scale parameters. However, the parameter related to \texttt{Europe} is significant to model the quantile of the mortality for COVID-19 only for small $q$. On the other hand, the four tests do not reject the normality assumption for the RQRs, suggesting that the RPGJSB1$_q$ model with the logistic distribution for $G$ and the cloglog link is appropriated to model all the considered quantiles of the mortality rate.

\begin{figure}[!h]
\psfrag{-7}[c][c]{\tiny{-7}}
\psfrag{-6}[c][c]{\tiny{}}
\psfrag{-5}[c][c]{\tiny{-5}}
\psfrag{-4}[c][c]{\tiny{-4}}
\psfrag{0.18}[c][c]{\tiny{0.18}}
\psfrag{0.16}[c][c]{\tiny{0.16}}
\psfrag{0.14}[c][c]{\tiny{}}
\psfrag{0.12}[c][c]{\tiny{}}
\psfrag{0.10}[c][c]{\tiny{0.10}}
\psfrag{0.08}[c][c]{\tiny{0.08}}
\psfrag{1.0}[c][c]{\tiny{1.0}}
\psfrag{0.5}[c][c]{\tiny{}}
\psfrag{0.0}[c][c]{\tiny{0.0}}
\psfrag{0.8}[c][c]{\tiny{0.8}}
\psfrag{0.6}[c][c]{\tiny{0.6}}
\psfrag{0.4}[c][c]{\tiny{0.4}}
\psfrag{0.2}[c][c]{\tiny{}}
\psfrag{-0.2}[c][c]{\tiny{-0.2}}
\psfrag{-0.4}[c][c]{\tiny{-0.4}}
\psfrag{1.2}[c][c]{\tiny{1.2}}
\psfrag{1.1}[c][c]{\tiny{1.1}}
\psfrag{1.0}[c][c]{\tiny{1.0}}
\psfrag{0.9}[c][c]{\tiny{}}
\psfrag{0.8}[c][c]{\tiny{0.8}}
\psfrag{0.7}[c][c]{\tiny{0.7}}
\psfrag{0.6}[c][c]{\tiny{0.6}}
\psfrag{0.2}[c][c]{\tiny{0.2}}
\psfrag{0.4}[c][c]{\tiny{0.4}}
\psfrag{0.6}[c][c]{\tiny{0.6}}
\psfrag{0.8}[c][c]{\tiny{0.8}}
\psfrag{1.0}[c][c]{\tiny{1.0}}
\psfrag{1.5}[c][c]{\tiny{1.5}}
\psfrag{1.2}[c][c]{\tiny{1.2}}
\psfrag{1.4}[c][c]{\tiny{1.4}}
\psfrag{1.6}[c][c]{\tiny{1.6}}
\psfrag{-2.5}[c][c]{\tiny{-2.5}}
\psfrag{-2.5}[c][c]{\tiny{-2.5}}
\psfrag{-2.5}[c][c]{\tiny{-2.5}}
\psfrag{55}[c][c]{\tiny{0.2}}
\psfrag{56}[c][c]{\tiny{0.4}}
\psfrag{57}[c][c]{\tiny{0.6}}
\psfrag{58}[c][c]{\tiny{0.8}}
\psfrag{CxC}[c][c]{\scriptsize{$q$}}
\begin{minipage}[b]{0.23\linewidth}
\psfrag{AxA}[c][c]{\scriptsize{$\beta_0(q)$}}
\centering
\includegraphics[width=4cm, angle=-90]{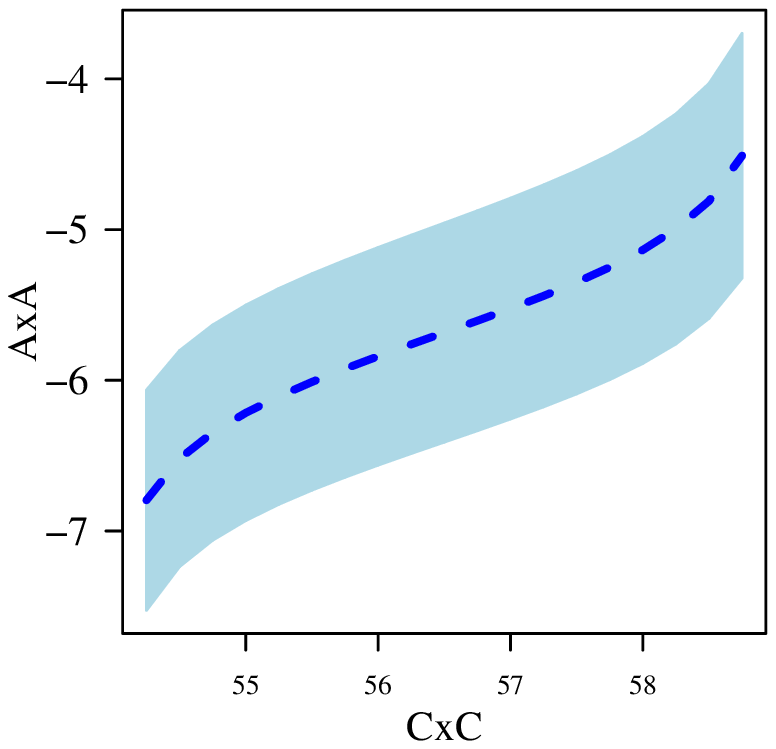}
\end{minipage} %\hfill
\hspace{0.2cm}
\begin{minipage}[b]{0.23\linewidth}
\psfrag{AxA}[c][c]{\scriptsize{$\beta_1(q)$}}
\centering
\includegraphics[width=4cm, angle=-90]{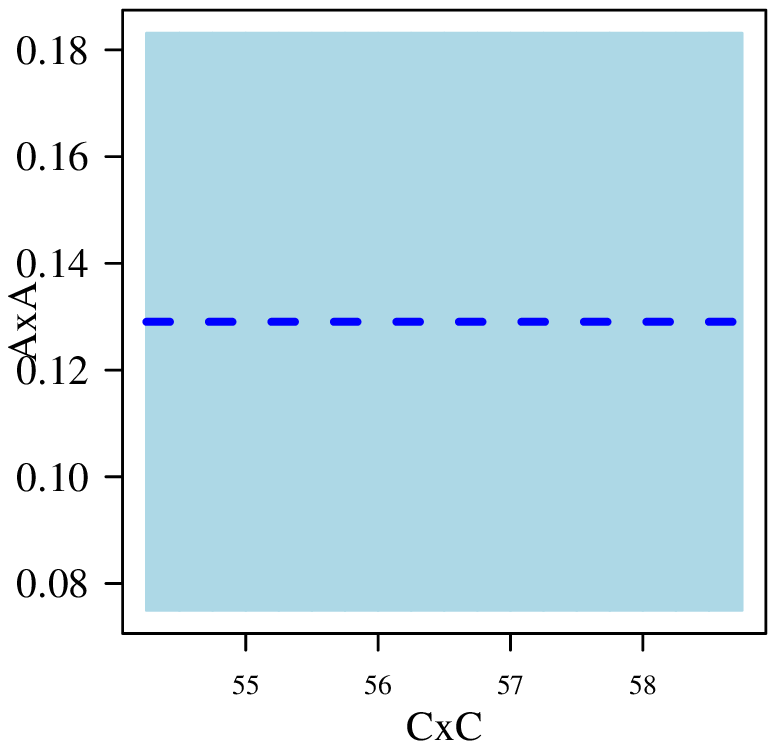}
\end{minipage}
\hspace{0.2cm}
\begin{minipage}[b]{0.23\linewidth}
\psfrag{AxA}[c][c]{\scriptsize{$\beta_2(q)$}}
\centering
\includegraphics[width=4cm, angle=-90]{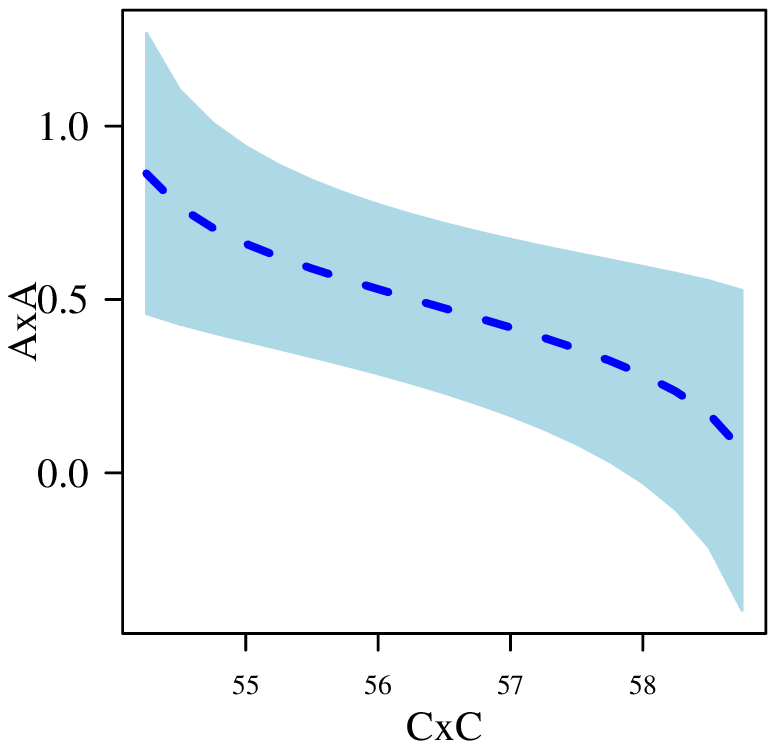}
\end{minipage} %\hfill
\hspace{0.2cm}
\begin{minipage}[b]{0.23\linewidth}
\psfrag{AxA}[c][c]{\scriptsize{$\beta_3(q)$}}
\psfrag{0.2}[c][c]{\tiny{}}
\psfrag{0.0}[c][c]{\tiny{}}
\centering
\includegraphics[width=4cm, angle=-90]{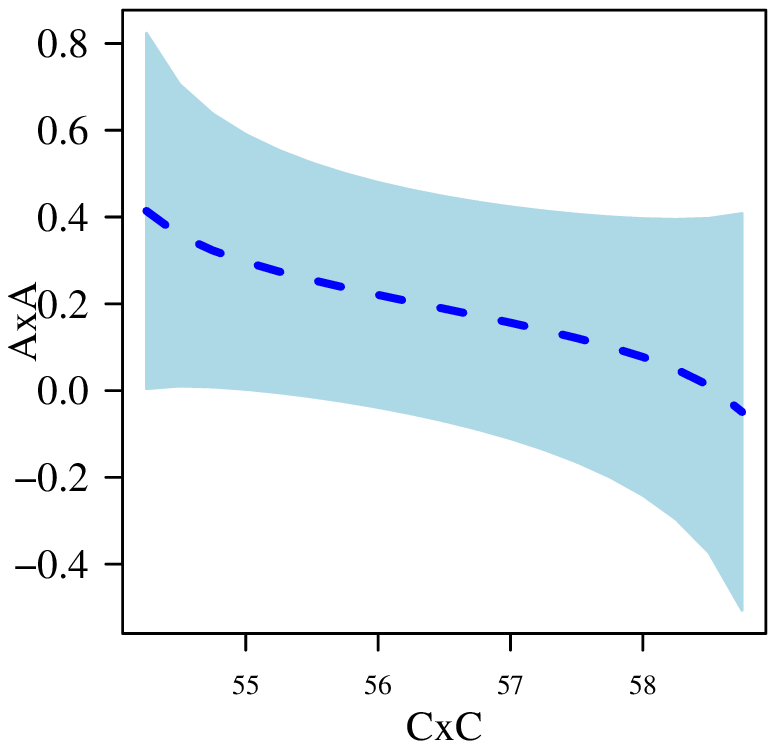}
\end{minipage} %\hfill
\\
%\begin{minipage}[b]{0.23\linewidth}
%\psfrag{AxA}[c][c]{\scriptsize{$\beta_4(q)$}}
%\centering
%\includegraphics[width=4cm, angle=-90]{figures/plot25}
%\end{minipage} %\hfill
%\hspace{0.2cm}
\begin{minipage}[b]{0.23\linewidth}
\psfrag{AxA}[c][c]{\scriptsize{$\nu_0(q)$}}
\psfrag{0.4}[c][c]{\tiny{}}
\centering
\includegraphics[width=4cm, angle=-90]{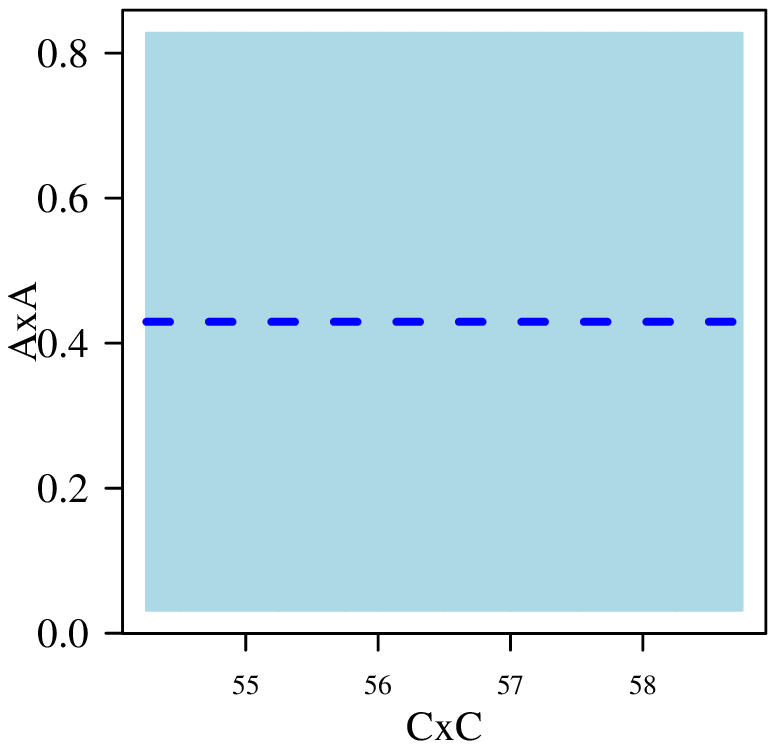}
\end{minipage}
\hspace{0.2cm}
\begin{minipage}[b]{0.23\linewidth}
\psfrag{AxA}[c][c]{\scriptsize{$\nu_1(q)$}}
\psfrag{0.5}[c][c]{\tiny{0.5}}
\psfrag{0.3}[c][c]{\tiny{}}
\psfrag{0.2}[c][c]{\tiny{}}
\centering
\includegraphics[width=4cm, angle=-90]{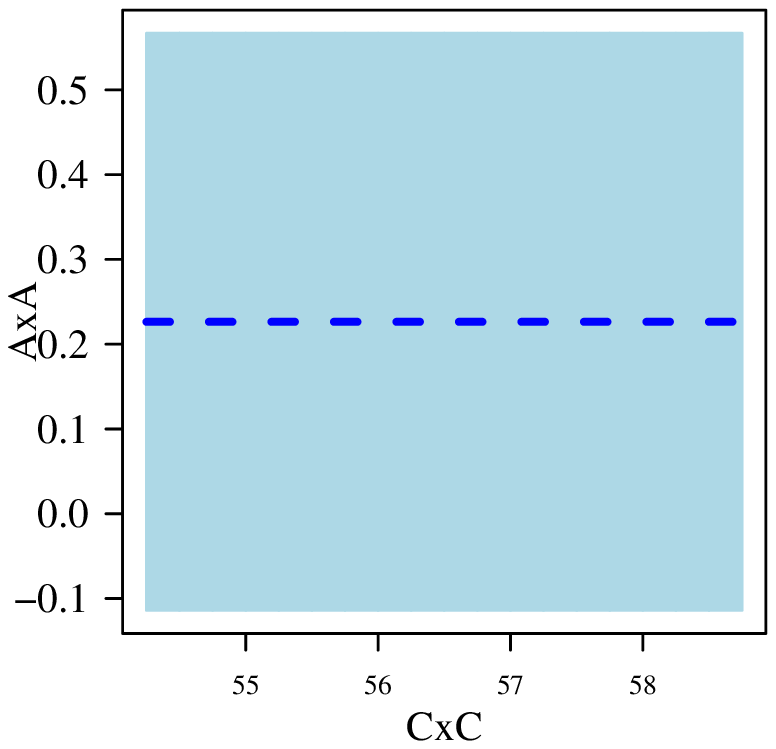}
\end{minipage} %\hfill
\hspace{0.2cm}
\begin{minipage}[b]{0.23\linewidth}
\psfrag{AxA}[c][c]{\scriptsize{$\nu_2(q)$}}
\psfrag{0.2}[c][c]{\tiny{}}
\psfrag{0.0}[c][c]{\tiny{}}
\centering
\includegraphics[width=4cm, angle=-90]{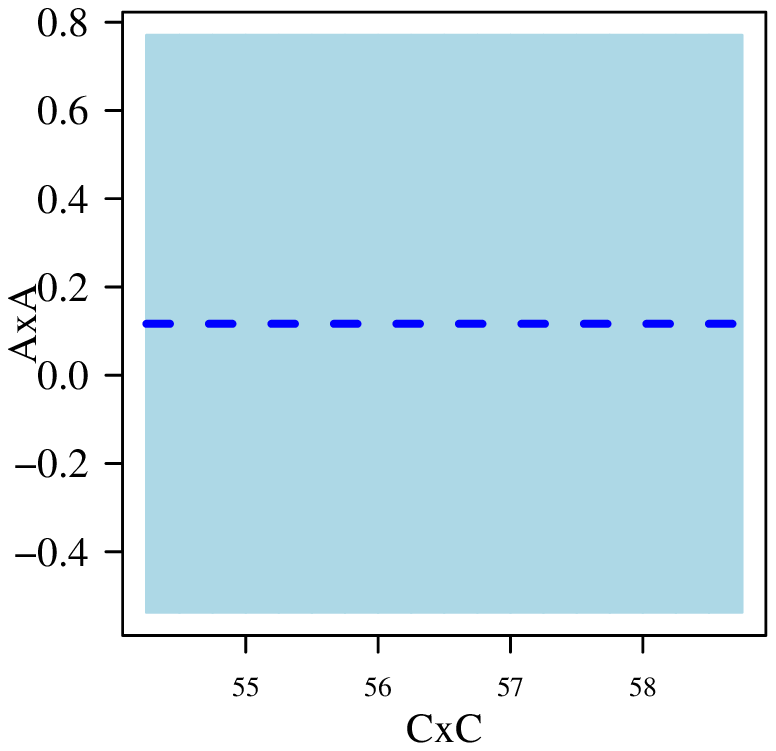}
\end{minipage} %\hfill
\caption{Point estimation and 95\% confidence interval for parameters estimated in RPGJSB1$_q$ model for different quantiles (cloglog link and G the cdf of the logistic model).}
\label{betas.q2}
\end{figure}

\begin{figure}[!h]
 \psfrag{AxA}[c][c]{\scriptsize{\texttt{mortality rate}}}
 \psfrag{CxC}[c][c]{\scriptsize{\texttt{log-surface (in km$^2$)}}}
 \psfrag{0.00}[c][c]{\tiny{0.00}}
 \psfrag{0.02}[c][c]{\tiny{0.02}}
 \psfrag{0.04}[c][c]{\tiny{0.04}}
 \psfrag{0.06}[c][c]{\tiny{0.06}}
 \psfrag{0.08}[c][c]{\tiny{0.08}}
 \psfrag{0.10}[c][c]{\tiny{0.10}}
 \psfrag{8}[c][c]{\scriptsize{8}}
 \psfrag{10}[c][c]{\scriptsize{10}}
 \psfrag{12}[c][c]{\scriptsize{12}}
 \psfrag{14}[c][c]{\scriptsize{14}}
 \psfrag{16}[c][c]{\scriptsize{16}}
 \psfrag{18}[c][c]{\scriptsize{18}}
 \psfrag{20}[c][c]{\scriptsize{20}}
 \centering
 \begin{minipage}[b]{0.28\linewidth}
\centering
\includegraphics[width=4.7cm]{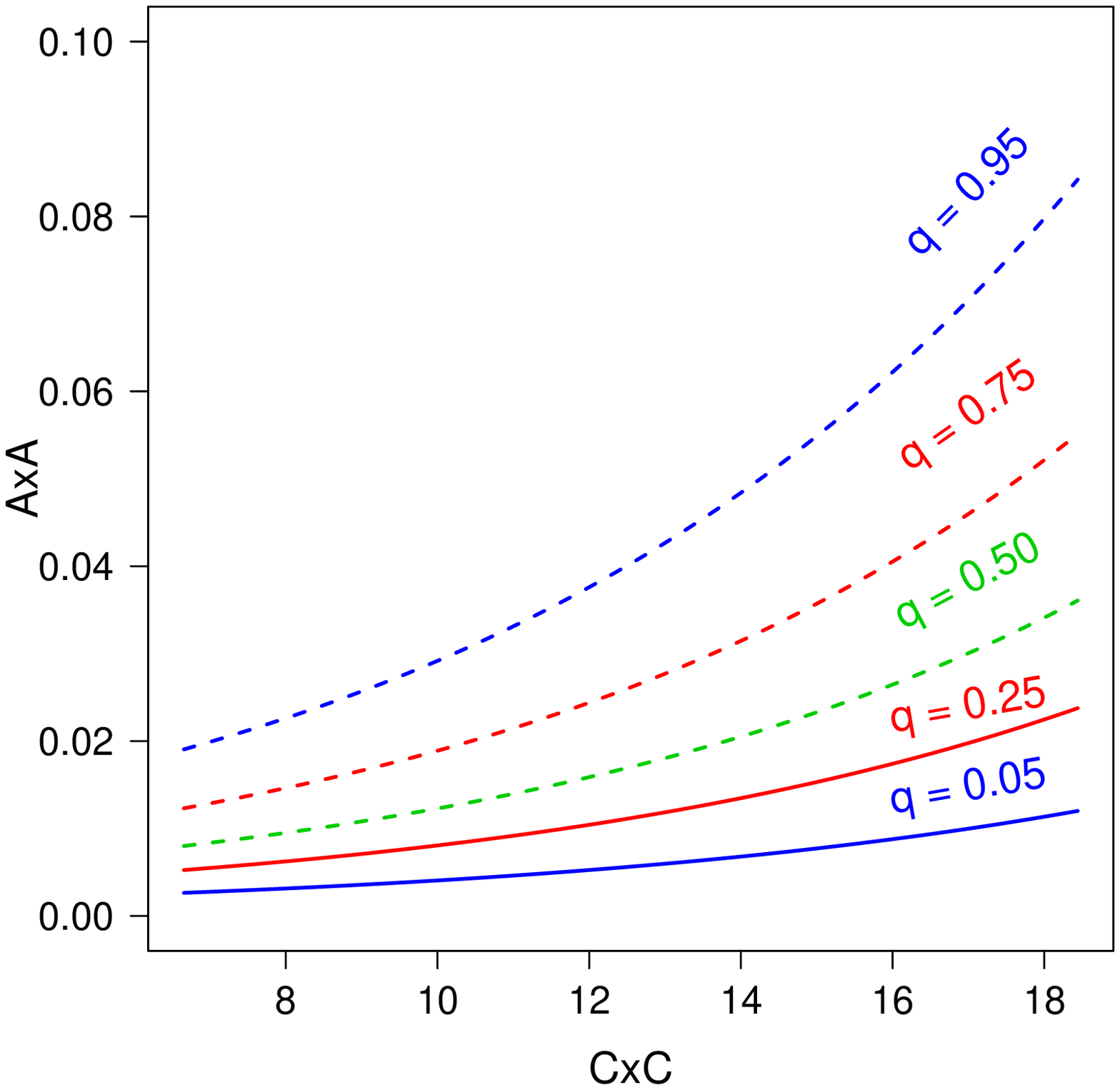}
\end{minipage} %\hfill
\hspace{0.3cm}
\begin{minipage}[b]{0.28\linewidth}
\centering
\includegraphics[width=4.7cm]{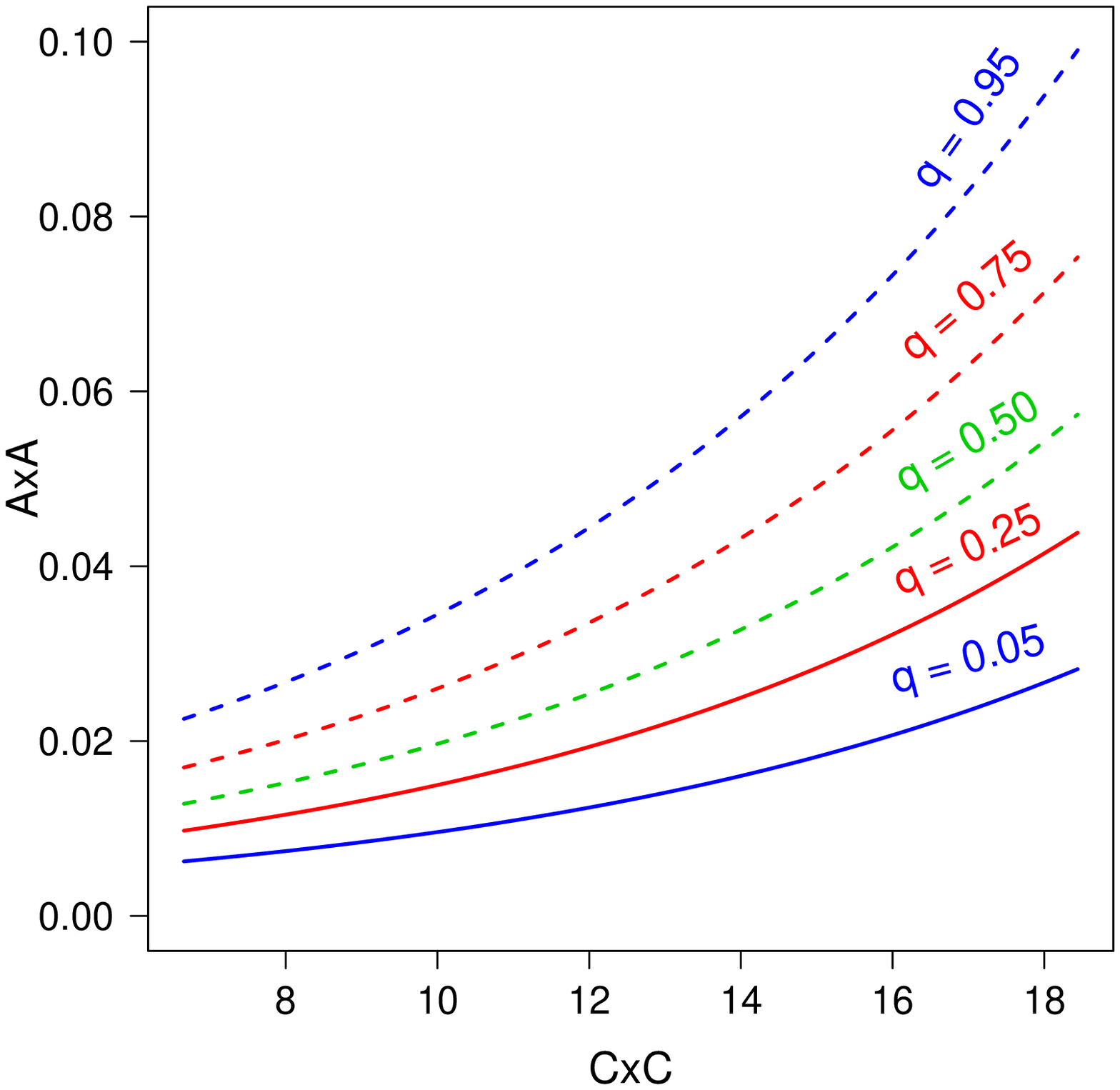}
\end{minipage}
\hspace{0.3cm}
\begin{minipage}[b]{0.28\linewidth}
\centering
\includegraphics[width=4.7cm]{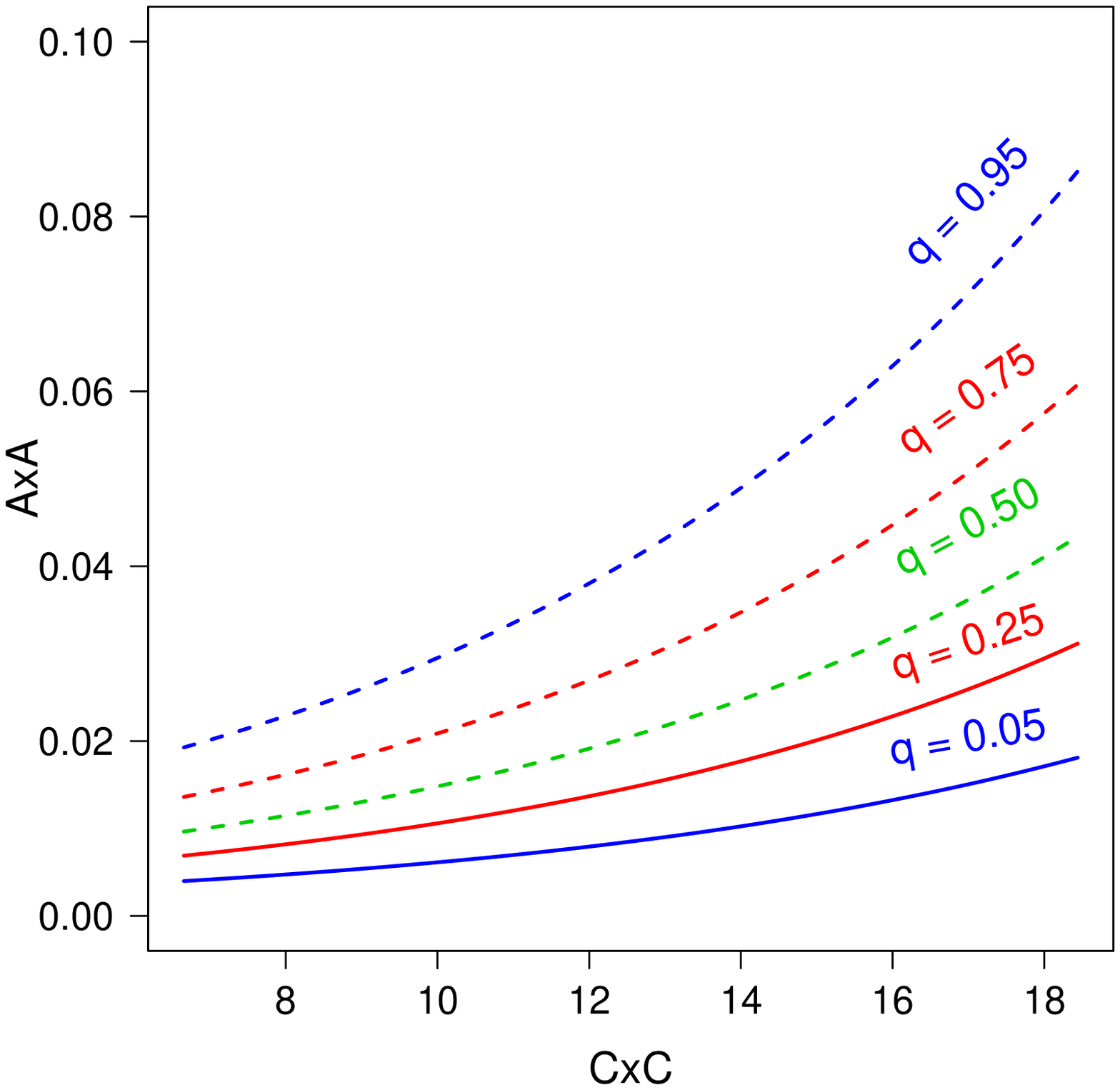}
\end{minipage}\\
 \caption{Estimated $100\times q$th quantile in the RPGJSB1$_q$ model varying the $\log(\texttt{surface})$ for countries in Africa, Asia or Oceania (left panel), America (center panel) and Europe (right panel) considering the cloglog link and $G$ the cdf from the logistic model.}
\label{estimated.quantiles2}
\end{figure}

On the other hand, Figure \ref{betas.q2} presented the point estimation and the 95\% confidence interval (CI) for the parameters in terms of the quantile $q$. From \ref{betas.q2}, the intercept for the quantile increases as $q$ increases, whereas the coefficients related to the quantile of \texttt{America} and \texttt{Europe} decreases when $q$ is increased. Furthermore, the coefficients related to the quantile for $\log(\texttt{surface})$ and the coefficients related to the scale of \texttt{America} and \texttt{Europe} remain similar for all $q$.
Figure \ref{estimated.quantiles2} presented the estimated quantiles $0.05, 0.25, 0.50, 0.75$ and $0.95$ for the mortality rate for different values of $\log(\texttt{surface})$.

\begin{table}[!h]
\begin{center}
\caption{Estimated parameters for different quantile in RPGJSB1$_{q=0.5}$ model for the COVID-19 data set with G the cdf of the logistic model and cloglog link. Also are presented the p-values for the traditional normality test for RQRs.}
\label{estimation.covid}
\resizebox{\linewidth}{!}{
\begin{tabular}{ccrcrrcccc}
\hline

 &          &            &            &            &            & \multicolumn{ 4}{c}{$p$-values for quantile residuals} \\

$q$ &     parameter      &  estimated &       s.e. &    $t$-value &    $p$-value &         KS &         SW &         AD &        CVM \\
\hline

           &      $\beta_0$ &    -5.6835 &     0.3709 &     -15.32 &    $<$0.0001 &            &            &            &            \\

           &      $\beta_1$ &     0.1290 &     0.0276 &       4.68 &    $<$0.0001 &            &            &            &            \\

           &      $\beta_2$ &     0.4749 &     0.1248 &       3.80 &     0.0001 &            &            &            &            \\

      0.50 &      $\beta_3$ &     0.1886 &     0.1320 &       1.43 &     0.0766 &      0.995 &      0.820 &      0.915 &      0.969 \\

           &        $\nu_0$ &     0.9060 &     0.1556 &       5.82 &    $<$0.0001 &            &            &            &            \\

           &        $\nu_1$ &     0.4294 &     0.2030 &       2.12 &     0.0172 &            &            &            &            \\

           &        $\nu_2$ &     0.2264 &     0.1737 &       1.30 &     0.0963 &            &            &            &            \\

           &        $\log \alpha$ &     0.1164 &     0.3337 &       0.35 &     0.3636 &            &            &            &            \\
\hline
\end{tabular}
}
\end{center}
\end{table}

\subsubsection{Local influence analysis}
\noindent

\begin{figure}[!h]
\centering
\psfrag{0.00}[c][c]{\scriptsize{0.00}}
\psfrag{0.02}[c][c]{\scriptsize{0.02}}
\psfrag{0.04}[c][c]{\scriptsize{0.04}}
\psfrag{0.06}[c][c]{\scriptsize{0.06}}
\psfrag{0.08}[c][c]{\scriptsize{0.08}}
\psfrag{0.10}[c][c]{\scriptsize{0.10}}
\psfrag{0}[c][c]{\scriptsize{0}}
\psfrag{20}[c][c]{\scriptsize{20}}
\psfrag{40}[c][c]{\scriptsize{40}}
\psfrag{60}[c][c]{\scriptsize{60}}
\psfrag{80}[c][c]{\scriptsize{80}}
\psfrag{100}[c][c]{\scriptsize{100}}
\psfrag{120}[c][c]{\scriptsize{120}}
\psfrag{4}[c][c]{\scriptsize{4}}
\psfrag{15}[c][c]{\scriptsize{15}}
\psfrag{49}[c][c]{\scriptsize{49}}
\psfrag{62}[c][c]{\scriptsize{62}}
\psfrag{80}[c][c]{\scriptsize{80}}
\psfrag{id}[c][c]{\scriptsize{index}}
\psfrag{CxA}[c][c]{\scriptsize{$C_{i}({\bm \beta})$}}
\psfrag{CxB}[c][c]{\scriptsize{$C_{i}({\bm \nu})$}}
{\includegraphics[height=5cm,width=5cm,angle=-90]{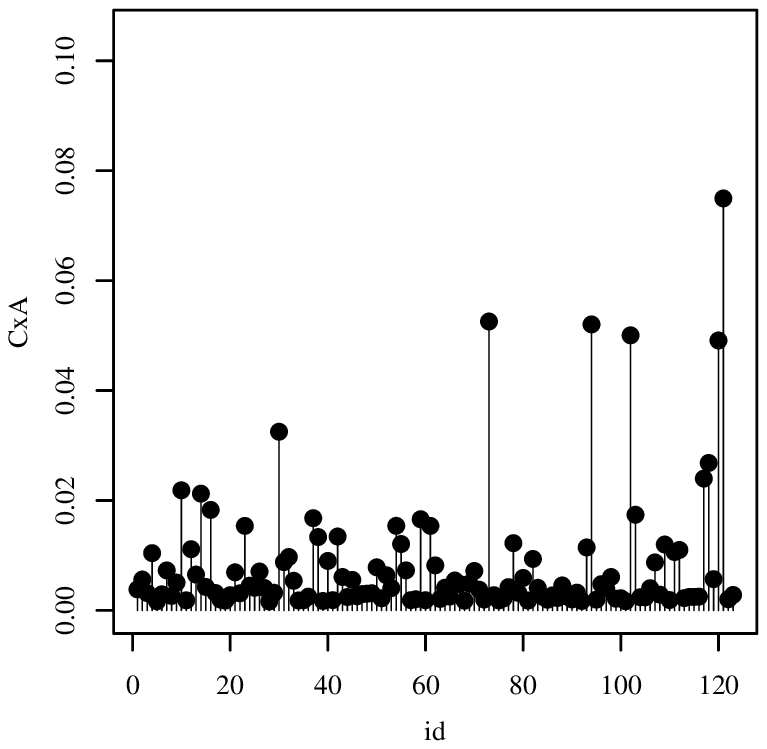}}
{\includegraphics[height=5cm,width=5cm,angle=-90]{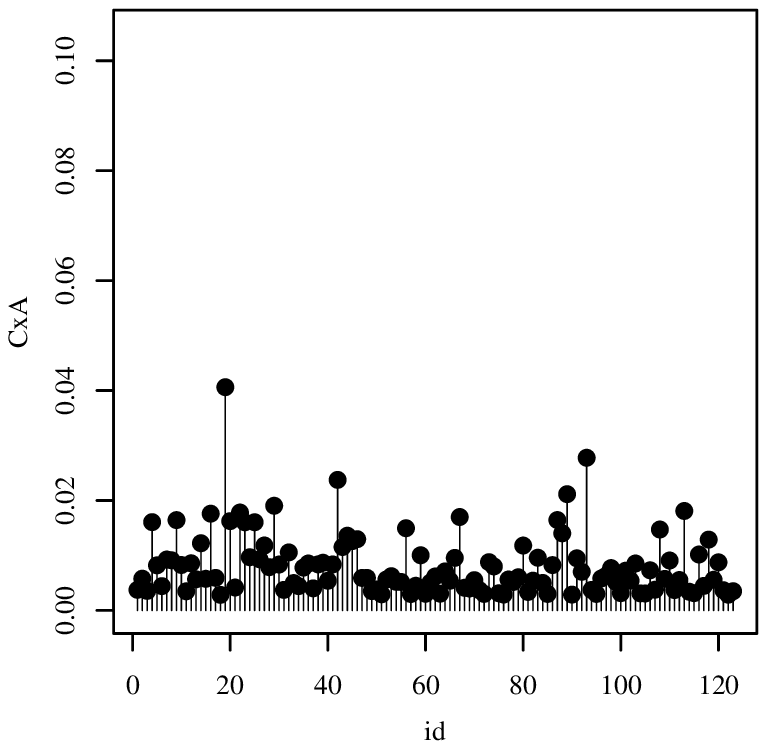}}
{\includegraphics[height=5cm,width=5cm,angle=-90]{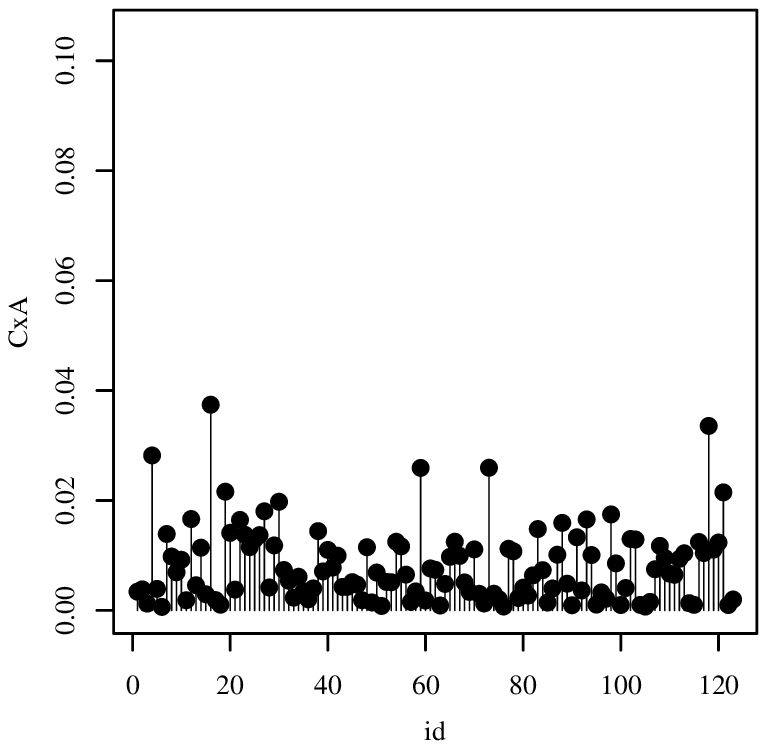}}
{\includegraphics[height=5cm,width=5cm,angle=-90]{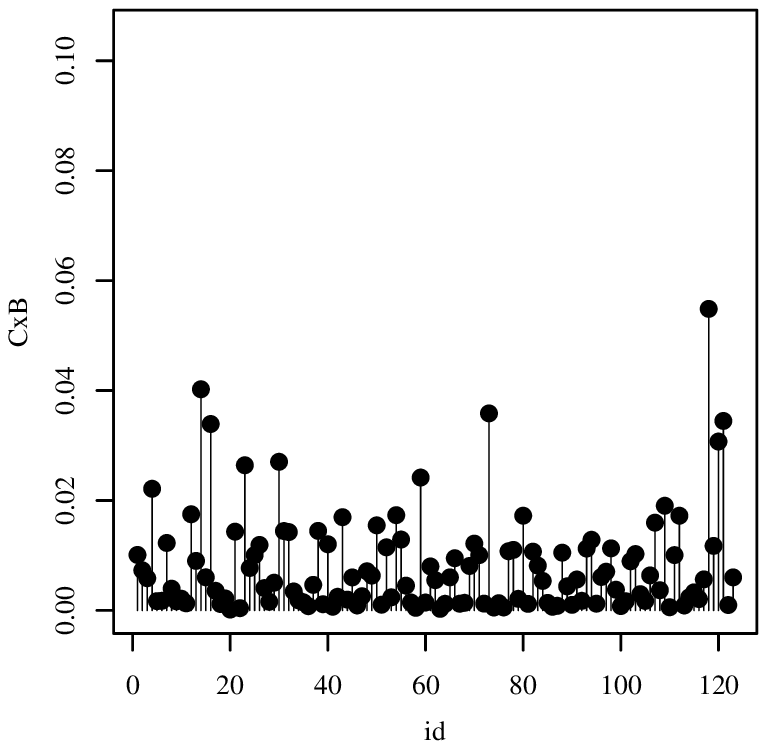}}
{\includegraphics[height=5cm,width=5cm,angle=-90]{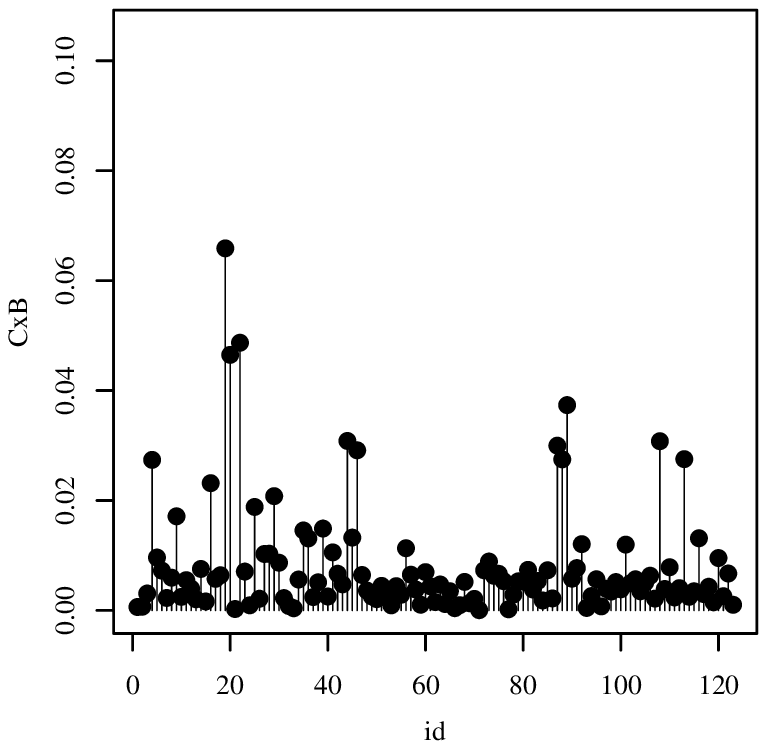}}
{\includegraphics[height=5cm,width=5cm,angle=-90]{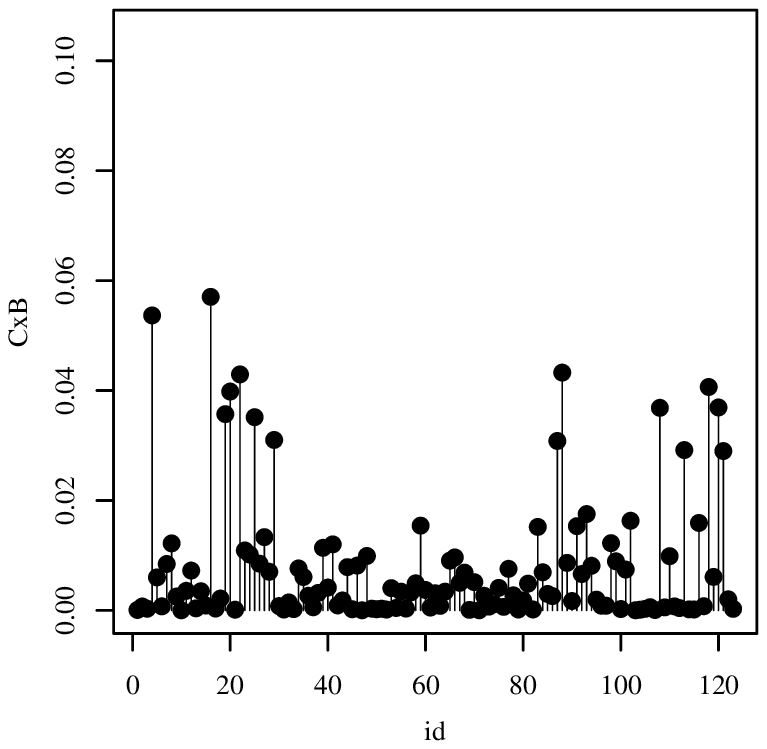}}
\caption{Index plots of $C_{i}$ for $\widehat{\bm \beta}$ (upper) and
$\widehat{\bm \nu}$ (lower) under the weight perturbation (left), response perturbation (center) and covariate perturbation (right) schemes for RPGJSB1$_{q=0.5}$ model (cloglog link and $G$ the cdf from the logistic model) in COVID-19 data set.}
 \label{fig:loc.inf.app2}
\end{figure}

We also presented a local influence analysis for the selected model under the three perturbations
schemes discussed in Section \ref{influence}. Figure \ref{fig:loc.inf.app2} shows such analysis for the RPGJSB1 model with
$q=0.5$ using the cloglog link and $G$ the cdf of the logistic model in the COVID-19 data set. In Section B.3 of the supplementary
material is presented the same analysis for other selected quantiles. Note that, considering all the cases,
the observation 121 appear in at least some case, which correspond to Yemen (Asia).
Yemen reported a high mortality rate (29\%, 601 accumulated deaths and 2067 accumulated cases, respectively). Evidently there is a problem in the handling of information about COVID-19 in the country.
Table \ref{estimation.withoutobsCOVID} presents
the relative change for the parameters (RC), for its estimated standard errors (RCSE) and the respective p-value
for the estimation without Yemen. We highlight that the greater variations are obtained for the parameters related to the scale and for $\log \alpha$ (excepting the case for $\beta_3(q=0.90)$). However, the estimated quantiles presented in Figure 3 do not depend on those parameters. Therefore, such plot without the referred observations are similar. We highlight that the significance of the parameters related to the quantile is maintained for all the cases (excepting for $\beta_3(q=0.05)$), suggesting a robustness of the model to estimate the different quantiles in this problem.

\begin{table}[!h]
\caption{RCs (in \%) in ML estimates and their corresponding SEs for the indicated parameter and
respective p-values for COVID-19 data set when observation 121 is dropped.}
\label{estimation.withoutobsCOVID}
\begin{center}
\begin{tabular}{ccrrrrr}
\hline
          &            &                                         \multicolumn{ 5}{c}{$q$} \\

           &  parameter &       0.10 &       0.25 &       0.50 &       0.75 &       0.90 \\
\hline

        RC &            &       7.81 &      11.06 &      16.43 &      23.58 &      32.58 \\

      RCSE &   $\beta_0(q)$ &       0.20 &       0.10 &       0.05 &       0.27 &       0.66 \\

   p-value &            &    $<$0.0001 &    $<$0.0001 &    $<$0.0001 &    $<$0.0001 &    $<$0.0001 \\
\cline{2-7}
        RC &            &      15.58 &      15.58 &      15.58 &      15.58 &      15.58 \\

      RCSE &   $\beta_1(q)$ &       0.28 &       0.28 &       0.28 &       0.28 &       0.28 \\

   p-value &            &    $<$0.0001 &    $<$0.0001 &    $<$0.0001 &    $<$0.0001 &    $<$0.0001 \\
\cline{2-7}

        RC &            &       1.17 &      13.45 &      41.56 &      99.46 &     266.08 \\

      RCSE &   $\beta_2(q)$ &       4.83 &       7.63 &       8.81 &       2.12 &      28.95 \\

   p-value &            &     0.0003 &     0.0001 &     0.0001 &    $<$0.0001 &    $<$0.0001 \\
\cline{2-7}

        RC &            &       9.43 &      30.59 &      78.19 &     216.11 &    2502.37 \\

      RCSE &   $\beta_3(q)$ &       6.92 &      12.61 &      16.00 &       1.10 &     450.43 \\

   p-value &            &     0.0526 &     0.0431 &     0.0383 &     0.0359 &     0.0351 \\
\cline{2-7}

        RC &            &      12.54 &      12.54 &      12.54 &      12.54 &      12.54 \\

      RCSE &   $\nu_0(q)$ &      28.63 &      28.63 &      28.63 &      28.63 &      28.63 \\

   p-value &            &     0.0562 &     0.0562 &     0.0562 &     0.0562 &     0.0562 \\
\cline{2-7}

        RC &            &       45.7 &       45.7 &       45.7 &       45.7 &       45.7 \\

      RCSE &     $\nu_1(q)$ &       8.75 &       8.75 &       8.75 &       8.75 &       8.75 \\

   p-value &            &     0.1588 &     0.1588 &     0.1588 &     0.1588 &     0.1588 \\
\cline{2-7}

        RC &            &      35.73 &      35.73 &      35.73 &      35.73 &      35.73 \\

      RCSE &     $\nu_2(q)$ &      11.86 &      11.86 &      11.86 &      11.86 &      11.86 \\

   p-value &            &     0.3219 &     0.3219 &     0.3219 &     0.3219 &     0.3219 \\
\cline{2-7}

        RC &            &     951.62 &     951.63 &     951.63 &     951.61 &     951.64 \\

      RCSE &     $\log \alpha(q)$ &     694.77 &     694.78 &     694.78 &     694.76 &     694.78 \\

   p-value &            &     0.3855 &     0.3855 &     0.3855 &     0.3855 &     0.3855 \\

\hline
\end{tabular}
\end{center}
\end{table}

\section{Conclusions}\label{sec:6}
\noindent

In this paper, we propose two classes of parametric quantile regression models for studying the association between a bounded response and covariates via inferring the conditional quantile of the response. The proposed quantile regression models was built based on power Johnson SB distribution \citep{Cancho} using a new parameterization of this distribution that is indexed by quantile, dispersion and shape parameters (RPGJSB1$_q(\psi,\delta,\alpha)$) or quantile and dispersion parameters (RPGJSB2$_q(\psi,\delta)$).
The first proposed quantile model has an extra-parameter $\alpha > 0$ is associated
with the ``tailedness'', and the second proposed quantile model has a less computational costs.
The ML inference was implemented to estimate the models
parameters, which was satisfactory considering the
simulation studies where parameters were recovered for different sample sizes.
Furthermore, under each proposed quantile regression model, we have developed model diagnostic tools.
In order to illustrate our approach, two applications using real data sets were presented and discussed. In particular, we analyze
the mortality rate of COVID-19 and the reproductive activity of the Chilean anchoveta. Results of the applications
showed that the proposed quantile models are adequate. Based on the results, the RPGJSB1$_q$ regression model presents a better fit for the COVID-19 mortality rate and the anchoveta data sets. As part of future research, there are several extensions of the new models not considered in this paper that can be addressed in future research, in particular, an extension of the methods developed in this paper would be to consider in (\ref{pdf.RPGJSB1}) a much more general family of distributions; that is, consider models for zero-inflated and one-inflated data set. Directions related to random effects in the model also can be addressed in future works.

\section*{Acknowledgements}
\noindent

The authors thank to ``Instituto de fomento pesquero'' (IFOP) to provide the anchoveta data set presented in the supplementary material.

\end{document}